\pdfoutput=1
\documentclass[preprint]{elsarticle}

\usepackage{float}
\usepackage{url}
\usepackage[export]{adjustbox}
\usepackage{wrapfig}

\usepackage{algorithm}
\usepackage[noend]{algpseudocode}
\usepackage{amssymb,amsmath,amsthm,amsfonts}
 \usepackage{booktabs}
 \usepackage{multirow}
\usepackage{subfig}
\newtheorem{definition}{Definition}

\usepackage{paralist}

\usepackage{graphicx}

\usepackage{color}
\usepackage[formats]{listings}
\lstdefineformat{C}{%
	\{=\newline\string\newline\indent,%
	\}=[;]\newline\noindent\string\newline,%
	\};=\newline\noindent\string\newline,%
	;=[\ ]\string\space}
\usepackage{listings}
\definecolor{mGreen}{rgb}{0,0.6,0}
\definecolor{mGray}{rgb}{0.5,0.5,0.5}
\definecolor{mPurple}{rgb}{0.58,0,0.82}
\definecolor{backgroundColour}{rgb}{0.95,0.95,0.92}
\usepackage[norndcorners,customcolors]{hf-tikz}
\definecolor{lightGrey}{rgb}{0.9, 0.9, 0.9}
\newcommand{\Hilight}{\makebox[0pt][l]{\color{lightGrey}\rule[-0.2em]{\linewidth}{1.0em}}}

\lstdefinestyle{CStyle}{
	commentstyle=\color{mGreen},
	keywordstyle=\color{magenta},
	stringstyle=\color{mPurple},
	basicstyle=\footnotesize,
	breakatwhitespace=false,         
	breaklines=true,                 
	captionpos=b,                    
	keepspaces=true,                 
	numbers=left,                    
	numbersep=5pt,                  
	showspaces=false,                
	showstringspaces=false,
	showtabs=false,                  
	tabsize=2,
	language=C
}

\biboptions{longnamesfirst,angle,semicolon}
\newcommand{\NAME}{\textbf{Twin-Finder+}}
\begin{document}
	\begin{frontmatter}
	\title{Integrated Reasoning Engine for Pointer-related Code Clone Detection}

\author{Hongfa Xue\corref{mycorrespondingauthor}}
\cortext[mycorrespondingauthor]{Corresponding author}
\ead{hongfaxue@gwu.edu}
\author{Yongsheng Mei }

\author{Kailash Gogineni }
 
\author{\\Guru Venkataramani}
\author{Tian Lan}
\address{The George Washington University}
\address{800 22nd st NW, Washington, DC}

	\begin{abstract}
		Detecting similar code fragments, usually referred to as \textit{code clones}, is an important task. In particular, code clone detection can have significant uses in the context of vulnerability discovery, refactoring and plagiarism detection. 
	    However, false positives are inevitable and always require manual reviews.
		In this paper, we propose \NAME, a novel closed-loop approach for pointer-related code clone detection that integrates machine learning and symbolic execution techniques to achieve precision. \NAME\ introduces a formal verification mechanism to automate such manual reviews process. Our experimental results show \NAME\ that can remove 91.69\% false positives in average. We further conduct security analysis for memory safety using real-world applications, Links version 2.14 and libreOffice-6.0.0.1. \NAME\ is able to find 6 unreported bugs in Links version 2.14 and one public patched bug in libreOffice-6.0.0.1. 
	\end{abstract}
\begin{keyword}
	Code Clone Detection\sep Machine Learning\sep Memory Safety

\end{keyword}
\end{frontmatter}



	%

	\section{Introduction}

Detecting similar code fragments, usually referred to as \textit{code clones}, is an important task, especially in large code bases~\cite{gabel2010study,kim2005empirical,li2006cp,xue2019machine}. Various software engineering tasks are taking advantage of code clone detection, such as vulnerability discovery, refactoring and plagiarism detection. Prior approaches have been proposed to detect code clones using token subsequence matching, tree or control flow based graph analysis~\cite{baker1997parameterized,kamiya2002ccfinder,jiang2007deckard}. However, they have limited scalability since the pairwise string or tree comparison is expensive in large code bases. On the other hand, machine learning-based clone detections are proposed to improve the previous string-matching based clone detections by introducing a code similarity measurement and transferring the code into intermediate representations (e.g. feature vectors) to detect more code clones~\cite{baxter2004dms,basit2005detecting,chen2017damgate,chen2018toss}. However, this may cause a considerable amount of false positives due to a smaller code similarity threshold.

In this paper, we introduce an integrated reasoning clone detection engine, \NAME, that is designed for better security analysis using code clone detection in large scale systems. Our approach uses domain-specific knowledge for code clone analysis, which can be used to detect code clone samples spanning non-contiguous and intertwined code base in software applications. As an example, since pointers and pointer-related operations widely exist in real-world applications and often cause security bugs~\cite{caballero2012undangle,serna2012info,conti2015losing}, detecting such pointer-specific code clones are of great significance. We note that a similar approach could be adopted to identify domains relating to any data-flow or control-flow specific code. 

In this work, we first perform pointer dependency analysis using light-\\weight tainting to traverse the program control flow graph and find pointer-related operations that can affect change buffer bounds. Then, we leverage both backward and forward program slicing to remove pointer irrelevant codes and isolate pointers in order to find non-contiguous and possibly intertwined pointer-related code clone samples. By doing so, we are able to improve the number of code clones detected, as well as the coverage of code base with respect to finding relevant code clones that ultimately helps with rapid security analysis. To facilitate higher code coverage, we also explore a wide range of code similarity threshold for the detection process. 

To verify the robustness of clone detection, we design a clone verification mechanism using symbolic execution (SE) that formally verifies if the two clone samples are indeed true code clones. We use a recursive sampling approach to randomly divide each grouped cluster into smaller ones. We sample each such smaller cluster of code clones and make all the pointer related variables as symbolic variables. We then apply symbolic execution to verify if they are true code clones, as SE is able to execute and explore all the possible paths to collect memory bound checking conditions (named as {\it constraints}). Two code clone samples are determined as true code clone pair if they both share the same memory safety constraints. Moreover, it is highly likely that code clone detection algorithm can still cause false positives. Existing works have reported that the false positives from code clone detection are inevitable~\cite{sajnani2016sourcerercc,allamanis2017survey} and human efforts are still needed for further verification and tuning detection algorithms. To automate this verification process, we introduce a feedback loop using formal analysis. We compare the Abstract Syntax Trees (AST) representing two code clone samples and add numerical weight to the feature vectors corresponding to the two code clone samples. We either decrease or increase the weights depending on the outputs of constraints comparision: (1) If two clone samples share the same constraints, it will be deemed as true code clone and we will decrease the weights; (2) if two clone samples share different constraints, which means they are false positives, we will increase the weights to eliminate such false positives. Finally, we exponentially recalculate the distances among feature vectors to reduce the false positives admitted from code clone detection. 

We have implemented a prototype of \NAME\ with two major modules: Domain Specific Slicing and Closed-loop Code Clone Detection. It utilizes several open-source tools and presents a new closed-loop operation with the assistance of formal analysis. In particular, we use a static code analysis tool, Joern~\cite{joern} and develop a program slicing framework. We instrument a tree-based code clone detection tool, DECKARD~\cite{jiang2007deckard}, to detect code clones after slicing. We employ a source code symbolic execution tool, KLEE~\cite{cadar2008klee}, for our clone verification and feedback to vector embedding in previous code clone detection module. 

We evaluate the effectiveness of \NAME\ in real-world applications, such as Links~\cite{links}, thttpd~\cite{thttpd_ACME}. For code clone detection, we apply \NAME\ to evaluate the number of code clones detected against conventional code clone detection approaches. We further construct security case studies for vulnerability discovery. The results show \NAME\ finds 6 unreported bugs in Links version 2.14 and one public reported bug in libreOffice-6.0.0.1, including 3 memory leaks and 3 Null Dereference vulnerabilities. And 1 of the memory leaks bug is silently patched in the newer version of Links. We further compare the overhead of our clone verification module using symbolic execution and the execution time with pure symbolic execution over entire binary programs to find the bugs.

The contributions of this paper are summarized as follows:

\begin{itemize}
	
	\item  We propose \NAME,  a pointer-related code clone detection framework intergated with static dode analysis and formal verification approach to detect non-contiguous and intertwined code clones.
	
    \item  \NAME\ introduces a clone verification mechanism to formally verify if two clone samples are indeed clones and a feedback loop to tune code clone detection algorithm and further reduce the false positives.
        
	\item We implement a prototype of \NAME\ using several open-source tools, including Joern, DECKARD, and KLEE. Our evaluation demonstrates that \NAME, with the optimal configuration, can detect up to 9$\times$ more code clones comparing to conventional code clone detection approaches and can remove 91.69\% false positives in average.
	\item  We conduct case studies of pointer analysis for memory safety using real-world applications 
	We show that using \NAME\ we find 6 unreported bugs in Links version 2.14 and one public patched bug in libreOffice-6.0.0.1. 

\end{itemize}

The rest of this paper is structured as follows: We first list the limitations of existing code clone detection approaches and the key solutions of our approach. 
We give the overview of \NAME\ in Section~\ref{ao} and introduce the designs and implemenations of \NAME\ along with technical details in Section~\ref{SD} and Section~\ref{Implementation}. We evaluate
its effectiveness to detect code clones in real-world applications and conduct a case study about security analysis in Section~\ref{eva}. Finally, we discuss the related work and concludes the paper in  Section~\ref{related} and Section~\ref{conclusion} respectively.
	\section{Problem Statement and Motivation}
\label{motivsation}
\begin{table}[]
		\centering

	\begin{tabular}{|c|c|c|c|}
		\hline
		\textbf{Similarity} & \textbf{\#True Positives} & \textbf{\#False Positives} & \textbf{\%False Positives} \\ \hline
		$=1.00$                & 1,495                     & 0                         & 0.00\%                     \\ \hline
		$\geq0.95$                & 2,016                     & 203                        & 9.15\%                     \\ \hline
		$\geq0.90$                & 2,637                     & 394                        & 13.00\%                    \\ \hline
		$\geq0.85$               & 3,017                     & 585                        & 16.24\%                    \\ \hline
		$\geq0.80$           & 3,526                     & 903                        & 20.75\%                    \\ \hline
	\end{tabular}
\caption{Clone statistics of true positives and false positives detected from sphinx3 benchmark using DECKARD}
\label{table:fp}
\end{table}

\begin{figure*}[h]
	\hspace{0.5cm}
\begin{minipage}[b]{0.44\linewidth}
\begin{lstlisting}[style=CStyle, frame=single,escapechar=\&]
void dict2pid_dump (...){
...
&\Hilight&for (i = 0; i < mdef->n_sseq; i++) {
&\Hilight&	fprintf (fp, "%5d ", i);
&\Hilight&	for (j = 0; j < mdef_n_emit_state(mdef); j++)
&\Hilight&		fprintf (fp, "%5d", mdef->sseq[i][j]);
..
}
..
} 
\end{lstlisting}
Code fragment of function \textit{sphinx3::dict2pid\_dump} as pointer $\{mdef->sseq\}$ are intertwined inside of the function
\end{minipage}
\hspace{0.5cm}
\begin{minipage}[b]{0.44\linewidth}	
\begin{lstlisting}[style=CStyle, frame=single,escapechar=\&]
int32 gc_compute_closest_cw (...){
...
&\Hilight&for(codeid=0; codeid< gs->n_code ;codeid+=2){
&\Hilight&	for(cid=0;cid<gs->n_featlen ; cid++)
&\Hilight&		fprintf (fp, "%5d", gs->codeword[codeid][cid]);
}
...
}
...
}
\end{lstlisting}
Code fragment of function \textit{sphinx3::gc\_compute\_closest\_cw} as pointer $\{gs->codeword\}$ are intertwined inside of the function
\vspace{0.04in}

\end{minipage}		
\caption{A true positive example}
\label{true_positives}
\end{figure*}
\begin{figure*}
\centering
\subfloat[][AST of function \textit{sphinx3::dict2pid\_dump} ]{\includegraphics[scale=0.25]{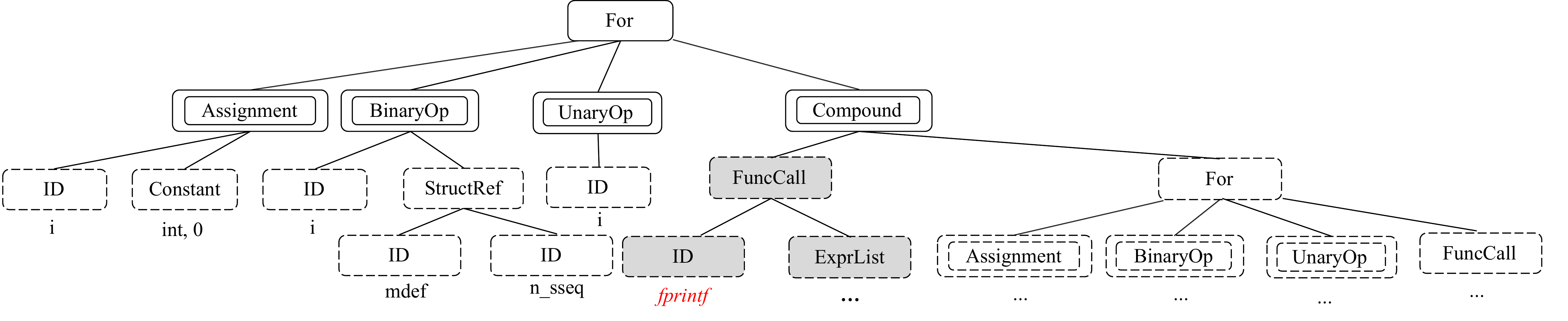}\label{<ast_3>}}
\hspace{0.2in}
\subfloat[][AST of function \textit{sphinx3::gc\_compute\_closest\_cw}]{\includegraphics[scale=0.25]{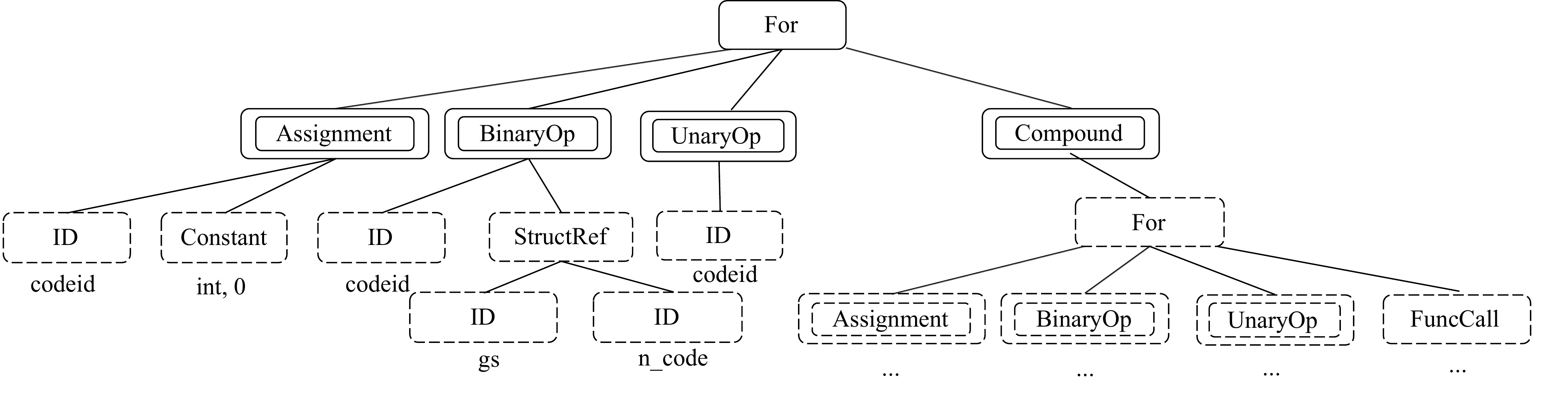}\label{<ast_4>}}

\caption{ASTs generated from the true positive example in Figure~\ref{true_positives}, where the shady nodes represent the different nodes between two trees}
\label{as_tp}
\end{figure*}

\begin{figure*}[h]
		\hspace{0.4cm}
\begin{minipage}[b]{0.44\linewidth}
	\begin{lstlisting}[style=CStyle, frame=single,escapechar=\&]
int32 mgau_eval (..., int32 *active)
{	
...
&\Hilight&for (j = 0; active[j] >= 0; j++) {
&\Hilight&	c = active[j];
	...
	}
...
}
\end{lstlisting}
\begin{lstlisting}[style=CStyle, frame=single,escapechar=\&]
void lextree_hmm_histbin (lextree_t *lextree,...)
{
...
&\Hilight&for (i = 0; i < lextree->n_active; i++) {
&\Hilight&	ln = list[i];
	..
	}
...
}
\end{lstlisting}
Code clone samples of function \textit{sphinx3::mgau\_eval} and \textit{sphinx3::lextree\_hmm\_histbin} as pointer $\{active\}$ and $\{list\}$ are intertwined inside of the functions
\label{example1}

\end{minipage}
\hspace{0.6cm}
\begin{minipage}[b]{0.44\linewidth}	
\begin{lstlisting}[style=CStyle, frame=single,escapechar=\&]
void fe_spec_magnitude(double *data, int32 data_len, double *spec, int32 fftsize)
{
...
IN = (complex *) calloc(fftsize,sizeof(complex));
...
&\Hilight&for (wrap=0; j<data_len; wrap++,j++) {
&\Hilight&	IN[wrap].r += data[j];
&\Hilight&	IN[wrap].i += 0.0;
&\Hilight&	}
	...
}
...
...
&\Hilight&for (j=0; j<fftsize;j++) {
&\Hilight&	IN[j].r = data[j];
&\Hilight&	IN[j].i = 0.0;
&\Hilight&	}
...
}
\end{lstlisting}

Code clone samples of function \textit{sphinx3::fe\_spec\_magnitude} as pointer $\{IN\}$ are intertwined inside of two different {\it for} loops of the function
\label{example2}

\end{minipage}		
\caption{False positive examples from sphinx3}
\label{false_positives1}
\end{figure*}
\begin{figure*}
	\centering
	\subfloat[][AST of function \textit{sphinx3::mgau\_eval} ]{\includegraphics[scale=0.25]{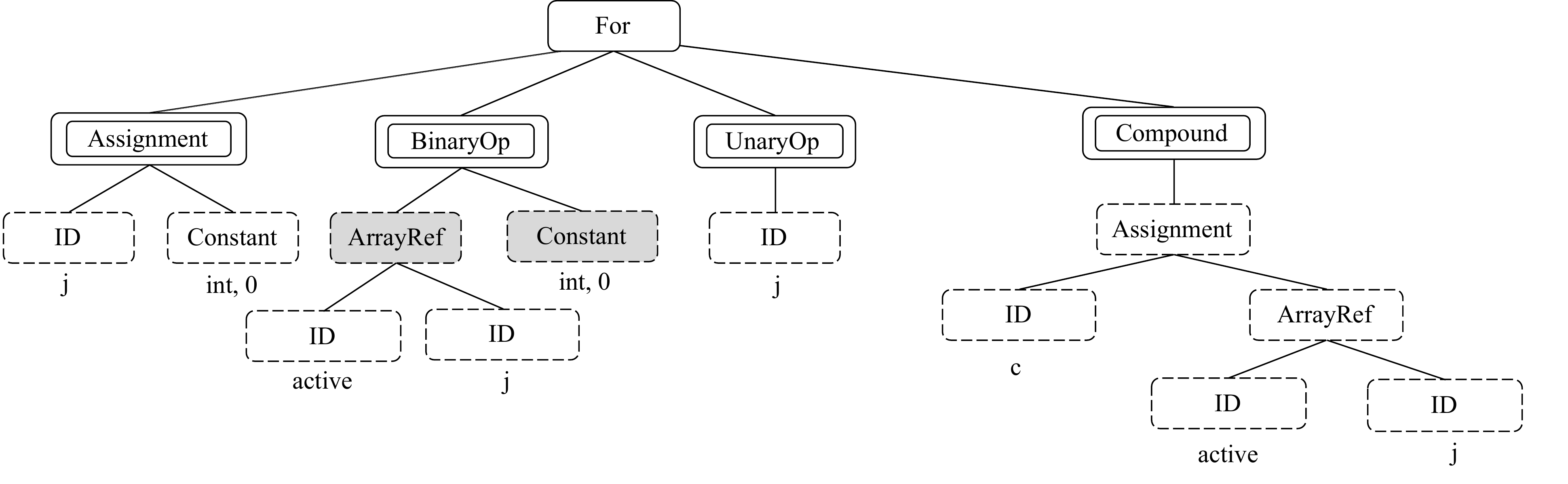}\label{<ast_1>}}
	\hspace{0.2in}
	\subfloat[][AST of function \textit{sphinx3::lextree\_hmm\_histbin}]{\includegraphics[scale=0.25]{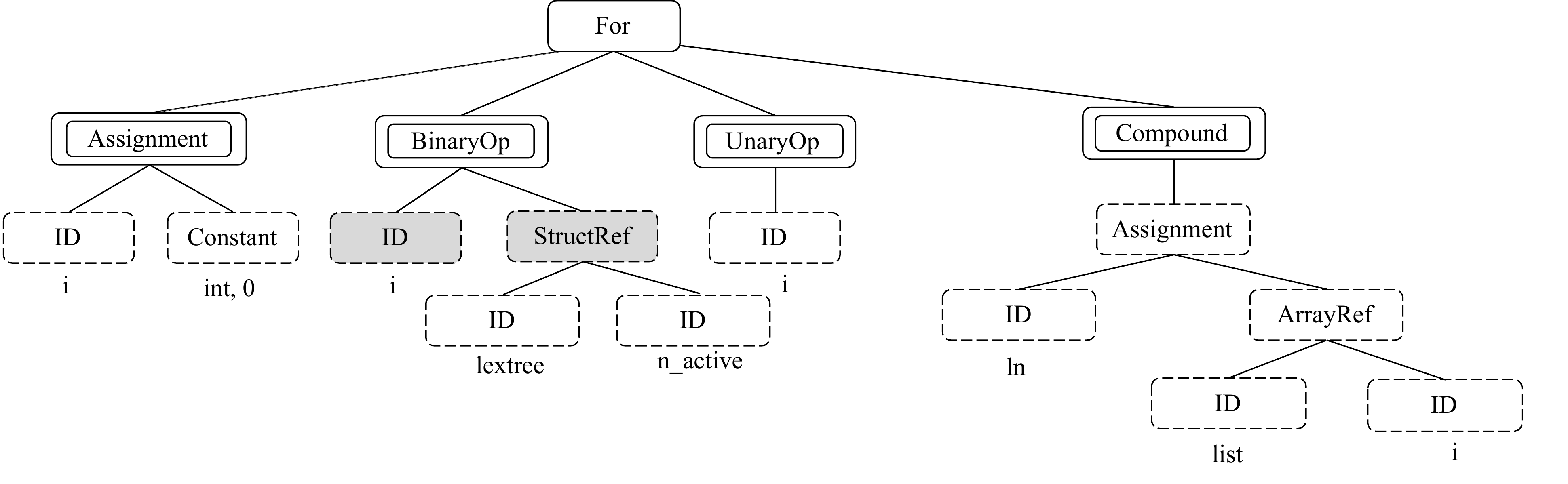}\label{<ast_2>}}
	
	\caption{ASTs generated from the firste false positive example in Figure~\ref{false_positives1}, where the shady nodes represent the different nodes between two trees}
	\label{as_fp}
\end{figure*}

\subsection{Code Clone Detection and Challenges}

Many software engineering tasks, such as refactoring, understanding code quality, or detecting bugs, require the
extraction of syntactically or semantically similar code fragments (usually referred to as {\it “code clones”}). Generally, there are three code clone types. \textbf{Type 1}: Identical code fragments except for variations in identifier
names and literal values; \textbf{Type 2}: Syntactically similar fragments that
differ at the statement level. The fragments have statements
added, modified, or removed with respect to each other. \textbf{Type 3}: Syntactically dissimilar code fragments that implement the same functionality.

Code clone detection approaches comprise two phases in general: (i) Transfer code into an intermediate representation, such as tree-based clone detection declaring feature vectors to represent code fragments~\cite{milea2014vector}; (ii) Deploy suitable similarity detection algorithms to detect code clones. For instance, clustering algorithms from machine learning are widely used in code clone detection problems~\cite{jiang2007deckard}. Some existing code clone detection techniques apply simple pattern matching (e.g., token-based code clone detection approach~\cite{baker1993program,kamiya2002ccfinder,li2006cp}) and leverage a code similarity metric to measure the amount of similarity between two code samples.   


In this paper, we aim to tackle two major issues from prior clone detection appraoches. 
	\begin{itemize}
		\item Pointers and pointer-related operations widely exist in real-world applications and often cause security bugs. Existing code clone detection approaches cannot detect only pointer-realted code clones, due to the considerable amount of pointer-irrelevant codes coupled with the target pointers
       \item Current clone detection approaches cannot guarantee zero false positives. Human efforts are always required for further verification. Here, we analyzed the true positives and the false positives detected using conventional tree-based code clone detection approach with different code similarity thresholds. We select \textit{sphinx3}, from SPEC2006 benchmark~\cite{spec}, as a representative application and the results are shown in Table~\ref{table:fp}. As we can see, relaxing code similarity threshold can benefit detection with more code clone samples. However, the ratio of false positives also increases at the time. If we can eliminate the false positives as many as possible, We still can enable a better analysis with more clone samples.

\end{itemize}

\subsection{Motivating Example}
\label{movtivating_example}
We select several false positive and true positive samples observed in sphinx3, detected from a tree-based clone detector DECKARD~\cite{jiang2007deckard}, as motiving examples.
First, we give the formal definition of false positive which is defined in Definition~\ref{fp}. 
\begin{definition}{\textbf{False Positives.}}
	\label{fp}
	In this paper, we define as false positives occur if a code clone pair is identified as code clones by code clone detection, but two clone samples share different bound safety constraints in terms of pointer analysis.
\end{definition}

Exisiting token or tree-based clonde detection always introduce a code similarity measurement $S$ and transfer the code into intermediate representations (e.g. feature vectors) to detect non-identical code clones. For example, the two code fragments shown in Figure~\ref{true_positives}. They can be further parsed and coverted into Abstract Syntax Trees (ASTs), where all identifier names and literal values are replaced by AST nodes. We genereate the ASTs for these two clone samples correspondingly in Figure~\ref{as_tp}. Both ASTs share a common tree pattern with only three different nodes appeared in the grey color nodes.

Assuming the target pointers for analysis are $mdef->sseq$ and $gs->codeword$, variables$\{i, j, mdef->n\_sseq, mdef\_n\_emit\_state(mdef)\}$ are identified as pointer-related variables (that can potentially affect the value of pointers) based on code dependency analysis. Thus, the bound safety conditions can be simply derived as these two equations. 
\begin{eqnarray}
\resizebox{.9\hsize}{!}
{$\{i < length(mdef->sseq)\} \land  \{j < length(*mdef->sseq)\}$}
\end{eqnarray} 
\begin{eqnarray}
\resizebox{.9\hsize}{!}
{$\{codeid < length(gs->codeword)\} \land  \{cid < length(*gs->codeword)\}$}
\end{eqnarray} 
respectively. As we can see, they are identical 
because the conditions differ only in variable names. Thus, they are true positives as they share the same pointer safety conditions. 

Even though a relaxed code similarity is able to detect such clones, it can also introduce a considerable amount of false positives.
Figure~\ref{false_positives1} illustrates two false-positive examples detected in sphinx3 from SPEC2006 benchmark. For the first example (showing on the left-hand side of the figure), two {\it for} loops are identified as code clones (line 4-5 and line 15-16) under a certain code similarity threshold. Figure~\ref{as_fp} shows the ASTs generated from those two code samples respectively. As we can see, they indeed share a common tree pattern but with 2 different nodes in shady color. Even though they are not identical, they still can be identified as similar looking code clones if we relax the code similarity threshold. Similarly, the second example (showing on the right-hand side) are sharing a similar code structure but differs only in identifier names. Thus, they can also be identified as code clones. 
Assuming the target pointers for analysis are $active$ and $list$ in the first example, we first to obtain pointer related variables through dependency analysis. It is easy to see that a solely variable $j$ is related to pointer $active$ but two variables $\{i, lextree->n\_active\}$ are related to $list$.
Thus, the bound safety conditions are deemed different. As mentioned in Definition~\ref{fp}, these two code clones will be defined as false positives since they do not share the same safety conditions. In the second example, the same dependency analysis procedure is deployed. Variables $\{wrap, j, data\_len \}$ are identified as pointer related variables in the first {\it for} loop (line 6-9) and $\{j, fftsize\}$ are related to second {\it for} loop (line 14-17), they are also false positives which are similar to the first example. One of the reasons to cause false positives in both cases are relaxed code similarity threshold to seek non-identical code clones. 

To formally verify if two code clones are true positives or false positives, symbolic execution can be applied to obtain memory safety conditions for further condition comparison. First, all pointer related variables of target pointers are made as symbolic variables. Symbolic execution can execute for each pointer dereference and generate array bounds safety conditions. To further eliminate false positives, in this paper, we propose a feedback loop to clone detection module through formal analysis. Once a false positive occur, we compare the ASTs representing two clone samples to find the different nodes and add numerical weight to those nodes so that we can recalculate the code similarity between two trees to reduce the false positives admitted from code clone detection. For example,  we note that the different nodes are \textbf{\{ID, StrucRef\}} and \textbf{\{ArrayRef, Constant\}} for the example showing in Figure~\ref{as_fp} respectively. Then we can simply add weight to each of those nodes. With a fixed code similarity, those two code samples will be eliminated in the future.

\section{Approach Overview}
\label{ao}

\begin{figure*}[h]

	\centering 
	\includegraphics[scale=0.5]{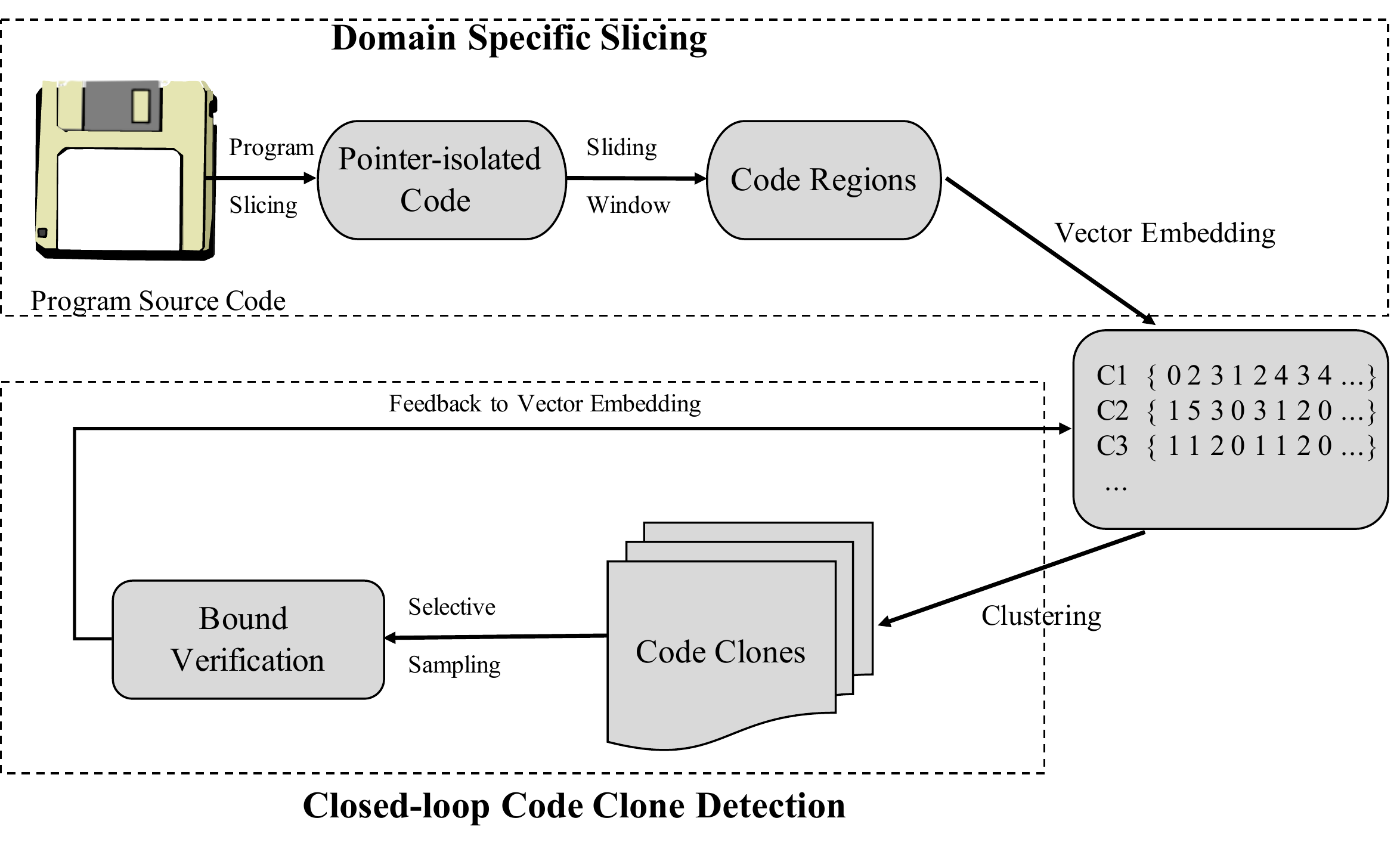}
	\caption{ Approach Overview}
	\label{fig:AO}
	
\end{figure*}
In this section, we give an overview of our framework. Two main components of \NAME\ are shown in Figure~\ref{fig:AO}, namely Domain Specific Slicing and Closed-loop Code Clone Detection. 

\textbf{Domain Specific Slicing}: We use tainting as a flexible mechanism to identify user-defined domains for further analysis. For a demonstration in this paper, we will use pointer analysis as the domain of interest. For a given program source code, Twin-Finder first generates dependency graph based on data and control information and uses a lightweight tainting approach to traverse the graph and find pointer related variables (such as array index). We then utilize program slicing to isolate pointers and the corresponding related statements. 

\textbf{Closed-loop Code Clone Detection}: After we generate pointer isolated code (containing pointers and their related code by including all of the variables and statements that affect them), code clone detection algorithm is applied to identify code clone samples. Toward this, we first generate Abstract Syntax Trees (AST) for each code fragment and transform such ASTs into feature vectors, embed them into vector space and use clustering algorithm from machine learning to find code clones. Note that we also use various code similarity thresholds to further increase the number of detected code clones(Section~\ref{clones}). 
A key distinguishing feature of our approach compared to prior work lies in improving the robustness of code clones detected through formal methods for verification. In particular, we use symbolic execution to verify whether the code clones grouped in the same clusters are indeed clones with respect to memory safety (that is, the pointer access is within legal array bounds). We define two clones are true clones only if they have the same array bound constraints (formed using pointer-affecting variables in the program code). We propose a recursive sampling mechanism for the clone verification and the process is performed as follows: In order to improve the analysis coverage among all the code clone pairs, we first randomly divide each cluster into smaller clusters. We then sample each such smaller cluster of code clones and apply symbolic execution to verify whether two clone samples share the same memory safety constraints derived using symbolic executors. We note that code clones within the same cluster may potentially have different constraints stemming from variables that affect their values. We consider such clones as \textit{false positives} introduced by code clone detection algorithm. If the selected clone samples are falsified by formal analysis, we enable a formal feedback mechanism to tune the feature vector weights accordingly to eliminate such false positives from occurring again. Section~\ref{verification} describes our design of this module.

	\section{System Design}
\label{SD}
In this section, we present details of our system design along with its modules.

\subsection{Program Slicing}\label{slicing}

We first name some certain types of variables as {\it pointer-related variables}. Those varibales can potentially affect the base, offset or bound information of a targeting pointer. For instance, pointer increment and array index are the most common {\it pointer-related variables}. 
In this paper, we use dependency analysis to find such pointer-related variables for each pointer on a function-level granularity. Then, we deploy both forward and backward program slicing to select related statements containing pointer and pointer-related variables.

\begin{enumerate}
	\item \textbf{Pointer Filtering.} 
	We first conduct static code analysis to collect all the pointers infromation from each function, including pointer declaration type (e.g. integer or string, local variables or global variables). A pointer list is then generated for each soruce program.
	In particular, we use a program parser ANTLR\cite{parr1995antlr} and a static code analysis tool Joren~\cite{joern} to analyze program syntax. 
	\item \textbf{Code Dependency Analysis and Tainting.} A directed dependency graph $\mathcal{DG}=(\mathcal{N}, \mathcal{E})$ is created for each pointer $p_i$ within the function where it is originally declared. The nodes of the graph $\mathcal{N}$ represent the identifiers in the function and edges $ \mathcal{E}$ represent the dependency between nodes, which reflects array indexing, assignments between identifiers and parameters of functions. As soon as the dependency graph is constructed, we start with the target pointer $p_i$ and traverse the dependency graph to discover all pointer-related variables in both top-down and bottom-up directions. This tainting propagation process stops at function boundaries. In the end, we generate the pointer-related variable list $p_i = \{v_1, v_2, \ldots , v_n\}$, where $v_i$ represents a pointer-related variable for pointer $p_i$. 
	\item \textbf{Program Isolation.} 
	After we obtain all the target pointers and their corresponding pointer-related variables,
	both forward and backward program slicing are deployed to isolate code into pointer-isolated code. Assuming a pointer-related variable list $V = \{v_1, v_2, \ldots, v_n\}$, we first take advantage of backward slicing: we construct a backward slice on each variable $v_i \in V$ at the end of the function and slice backwards to only add the statements into slice iff there is data dependency as $v_i $  is on left-hand side of assignments or parameter of functions, which can potentially affect the value of $v_i$, in the slice. 
	For example, a line of statement $v_i = x$ will be kept, but $y=v_i$ will be removed since it cannot change the value of $v_i$. 
	Whenever $v_i$ is in a loop (e.g. $while/for$ loop) or $if-else$/$switch$ branches, forward slicing is then used to add those control dependency statements to the slice. After performing program slicing, we are able to isolate one single function into several pointer separated functions. For instance, if there are 10 pointers in one function, then there should be 10 pointer isolated functions derived from this function. Note that it is possible that one statement involves multiple pointers, this type of statements will be selected in all the involved pointers. In additions, we also need to preserve the locations (e.g. line of code ) of any selected statements in the original source code for further analysis.
\end{enumerate} 

\subsection{Code Clone Detection}
\label{clones}
\NAME\ leverages a tree-based code clone detection approach, which is originally proposed by Jiang et al.~\cite{jiang2007deckard}. It produces the Abstract Syntax Tree (AST) representation of the source program to detect code clones by comparing subtrees in ASTs with a specific similarity metric. AST is commonly used tree representation by compilers to abstract syntactic structure of the code and to analyze the dependencies between variables and statements. The source code can be parsed by using the static code analysis mentioned in Section~\ref{slicing} and generate AST correspondingly. Here, we adopt the notions of code similarity, feature vectors and other related definitions from previous works~\cite{baxter1998clone,jiang2007deckard}. We deploy such method on the top of our domain specific slicing module to only detect code clones among pointer isolated codes. 
 
\subsubsection{Definitions}
We first formally give the several definitions used in our code clone detection module. 
\begin{definition}{\textbf{Code Similarity.}}

	Given two Abstract Syntax Trees (AST) $T_1$ and $T_2$, which are representing two code fragments, the weighted code similarity $S$ between them is defined as:
	\begin{eqnarray}
		 S (T_1, T_2) = \frac{2S}{2S+L+R}
		\end{eqnarray} \label{similarity}
$S$ is the number of shared nodes in $T_1$ and $T_2$; $\{L:[t_1, t_2, .., t_n],R:[t_1, t_2, .., t_m]\}$ are the different nodes between two trees, where $t_i$ represents a single AST node.
\end{definition}
\begin{definition}{\textbf{Feature Vectors.}}
	A feature vector $V = (v_1, v_2,...,v_n)$ in the Euclidean space is generated from a sub-AST, corresponding to a code fragment, where each $v_i$ represents a specific type of AST nodes and is calculated by counting the occurrences of corresponding AST node types in the sub-AST. More details related to AST nodes types can be found in~\cite{basit2005detecting}.
\end{definition}
Given an AST tree $T$ , we perform a post-order traversal of $T$ to generate vectors for its subtrees. Vectors
	for a subtree are summed up from its constituent subtrees.
{\it Example.} The feature vector for the code fragment of function \textit{sphinx3::mgau\_eval}, mentioned in Section~\ref{movtivating_example}, is $<7,2,2,2,0,1,1,1,1>$ where the ordered dimensions of vectors are
occurrence counts of the relevant nodes: \textbf{ID, Constant, ArrayRef, Assignment, StrucRef, BinaryOp, UnaryOp, Compound,} and \textbf{For}.
\subsubsection{Clone Detection}
Given a group of feature vectors, we utilize Locality Sensitive Hashing (LSH)\\~\cite{datar2004locality} and near-neighbor querying algorithm based on the euclidean distance between two vectors to cluster a vector group, where LSH can hash two similar vectors to the same hash value and helps near-neighbor querying algorithm to form clusters~\cite{jiang2007deckard,gionis1999similarity}. Suppose two feature vectors $V_i$ and $V_j$ representing two code fragments $C_i$ and $C_j$ respectively. The code size (the total number of AST nodes) are denoted as $S(C_i)$ and $S(C_j)$. The euclidean distance $E ([V_i ; V_j])$ and hamming distance $H ([V_i ; V_j] )$ between $V_i$ and $V_j$ are calculated as following:
\begin{eqnarray}
\label{distance}
E ([V_i ; V_j] )= ||V_i - V_j||_2^2
\end{eqnarray}
\begin{eqnarray}
\label{hamming}
H ([V_i ; V_j] )= ||V_i - V_j||_1
\end{eqnarray}
The threshold used for clustering can be approximated using the euclidean distance and hamming distance between two feature vectors for two ASTs $T_1$ and $T_2$ as following:
\begin{eqnarray}
\label{threshold}
E ([V_i ; V_j] ) \geq \sqrt{H ([V_i ; V_j] )} \approx \sqrt{L+R}
\end{eqnarray}
Based on the definition from Equation~\ref{similarity}, we can derive that \\$\sqrt{L+R} = \sqrt{2 (1-S) \times (|T_1|+|T_2|)}$, where  $(|T_1|+|T_2|) \geq 2 \times min (S(C_i), S(C_j))$. Then, the threshold for the clustering procedure is defined as:
\begin{eqnarray}
T = \sqrt{2 (1-S) \times min (S(C_i), S(C_j))}
\end{eqnarray}
Then, given a feature vector group $V$, the threshold can be simplified as $2(1 - S) \times min_{v \in V} \in S(v)$, where we use vector sizes
to approximate tree sizes.
The $S$ is the code similarity metric defined from Equation~\ref{similarity}. Thus, code fragments $C_i$ and $C_j$ will be clustered together as code clones under a given code similarity $S$ if $E ([V_i ; V_j] ) \leq T$.
\subsection{Bound Verification}
\label{verification}

To {\it formally check} if the code clones detected by \NAME\ are indeed code clones in terms of pointer memory safety, we propose a clone verification mechanism and utilize symbolic execution as our verification tool. 

There are three phases of clone verification: (1) Recursive sampling code clones in clusters;(2) Deploy symbolic execution and constraints solving for clone verification; (3) A feedback mechanism to vector embedding in previous code clone detection module to improve the correctness of clustering algorithm and eliminate false positives.

\subsubsection{Recursive Sampling}
To improve the coverage of code clone samples in the clusters, we propose a recursive sampling procedure to select clone samples for clone verification.

First, we randomly divide one cluster into several smaller clusters. Then we pick random code clone samples from each smaller cluster center and cluster boundary. After, we employ symbolic execution in selected samples for further clone verification. Note that the code clone samples are pointer isolated code generated from program slicing. Since symbolic execution requires the code completeness, we map the code clone samples to the original source code locations to perform partial symbolic execution.

\subsubsection{Clone Verification}
\label{clone_verification}
Clustering algorithm cannot offer any guarantees in terms of ensuring safe pointer access from all detected code clones. It is possible that two code fragments are clustered together, but have different bound safety conditions, especially if we use a smaller code similarity. To further improve the clone detection accuracy of Twin-finder we design a clone verification method to check whether the code clone samples are true clones.

Let $X=\{p_1,p_2,...,p_n\}$ be
a finite set of pointer-related variables as symbolic variables, while symbolic executing a program all possible paths, each path maintains a set of \textit{constraints} called the {\it path conditions} which must hold on the execution of that path.
First, we define an atomic condition, $AC()$,
over $X$ is in the form of $f(p_1,p_2,...,p_n)$, where
$f$ is a function that performs the integer operations
on $O \in \{>,<,\geq,	\le,=\}$. Similarly, a condition over $X$ can be a Boolean combination of
path conditions over $X$.
\begin{definition}{\textbf{Constraints.}}
	\label{constraints}
	An execution path can be represented as a sequence of basic blocks. 
	Thus, path conditions can be computed as $AC(b_0) \wedge AC(b_1) ...\wedge AC(b_n)$ where each $AC(b_i)$ in $AC()$ represents a sequence of atomic condition in the basic block $b_n$. For the case of involving multiple execution paths, the final constraints will be the union of all path conditions.
\end{definition}

Give a clone pair sampled from the previous step, we perform symbolic execution from beginning to the end of clone samples in original source code based on the locations information (line numbers of code). The symbolic executor is used to explore all the possible paths existing in
the code fragment. We collect all the possible constraints(defined in Definition~\ref{constraints}) for each clone sample after symbolic execution is terminated. 
To deal with possibly incomplete program state while performing partial symbolic execution, we only make the pointer-related variables in such code fragment as symbolic variables. We collect all the possible constraints(defined in Definition~\ref{constraints}) for each clone sample after symbolic execution is terminated. 
Then the verification process is straightforward. We deploy pairwise comparison of constraints between two clone samples. Two code clones are verified as true code clone if the constraints are an exact match. Since we analyze pointers in our work, constraints do not overlap in terms of the memory reference. However, it is possible for symbolic execution to generate more than one set of constraints since the clone samples may have multiple paths. In this case, we need to combine and format all the constraints into one. There are some exiting tools which can be used to solve and combine such constraints~\cite{zheng2013z3,brummayer2009boolector,veanes2010symbolic}. 
Then the verification process is straightforward. A constraint solver can be used to check the satisfiability and syntactic equivalence of logical formulas over one or more theories.

The steps of this verification process are summarized as follows:
\begin{itemize}
	\item \textbf {Matching the Variables}: To verify if two sets of constraints are equal, we omit the difference of variable names. However, we need to match the variables between two constraints based on their dependency of target pointers.
	For instance, two pointer dereference $ a[i] = 'A'$ and $b[j] = 'B'$, the indexing variables are $i$ and $j$ respectively. During symbolic execution, they both will be replaced as symbolic variables, and we do not care much about the variables names. Thus, we can derive a precondition that $i$ is equivalent to $j$ for further analysis.
	This prior knowledge can be easily obtained through dependency analysis mentioned in Section 4. 
	\item \textbf{Simplification}: Given a memory safety condition $S$, it can contain multiple linear inequalities. For simplicity, the first step is to find possibly simpler expression $S'$, which is equivalent to $S$.
	
	\item \textbf{Checking the Equivalence}: To prove two sets of constraints $S_1 == S_2$ ,we only need to prove the negation of $S_1 == S_2$ is unsatisfiable.

\end{itemize}
{\it Example.} Assuming we have two sets of constraints, $S_1 = (y_1\geq 10) \wedge(x_1 \geq 20)$ and $S_2 = (y_2\geq 10) \wedge(x_2 \geq 20)$, where $y_1$ is equivalent to $y_2$  and $x_1$ is equivalent to $x_2$. We then can solve that $Not(S_1 == S_2)$ is unsatisfiable. Thus, $S_1 == S_2$.

\subsection{Formal Feedback to Vector Embedding}\label{feedback}

\begin{algorithm}[h]
	\caption{Algorithm for Feedback to Vector Embedding}
	\label{alg1}
	\begin{algorithmic}[1]
		
		\State \text{\textbf{Input:}: Code Clone Samples $C_i$, $C_j$} $  $
		\State Corresponding AST sub-trees: $S_i $, $S_j $
		\State 	Corresponding Feature Vectors: $V_i$, $V_j$ 
		\State 	Current Code similarity threshold: $S$ 
		\State  Longest Common Subsequence \textbf{function:} LCS ()
		\State \text{\textbf{Output:}: Optimized Feature vectors: $O_i$, $O_j$}
		\State  \textbf{Initialization:}
		\State$O_i, O_j = V_i, V_j$
		\State $D = LCS(S_i, S_j)$
		\If{$C_i$ and $C_j$ share same constraints}

		\State $S_i = RemoveSubtrees(S_i - D)$ 
		
		\State $S_j = RemoveSubtrees(S_j - D)$
		
		\State	$O_{n \in \{i;j\}} =Vectornize(S_{n \in \{i;j\}})$;

		\Else
		\State T=[]
		%
		%
		\State $Uncommon\_Subtrees = (S_i - D) + (S_j - D)$
		\State $T.append(Uncommon\_Subtrees)$
		\For{$t$ in $T$}
		\If{$EuclideanDistance(O_i, O_j) < \sqrt{2(1 - S) \times min\{Size(V_i),Size(V_j)\} } $}
		\State$break;$
		\EndIf
		\State	$ t = d.index$\;  
		\State	$O_{n \in \{i;j\}}[t] = O_{n \in \{i;j\}}[t]*\delta$; \it where $\delta > 1.0$
		
		\EndFor	
		
		\EndIf
		
	\end{algorithmic}
\end{algorithm}

Now we describe a feedback mechanism to vector embedding in code clone detection if we observe false positives verified through the execution in Section~\ref{clone_verification}. 

A feedback process is a algorithm we propose in this paper to reduce false positives by tuning the feature vectors weights to the vector embeddings. The general idea of our feedback is that we analyze the difference between two ASTs by comparing two trees and find the differences in between. Then we add numerical weights to the feature vectors of two code clones to either increase or decrease the distance between them based on the outputs from the clone verification step. Once the weight is added, we re-execute the clustering algorithm in code clone detection module over the same code similarity threshold configuration. Note that this procedure can be executed in many iterations as long as we observe false positives from clone verification step. Furthermore, we can expect that such false positives are eliminated due to unsatisfied vector distance and out of cluster boundary.

Algorithm~\ref{alg1} shows the steps of feedback in detail. Given a code similarity threshold $S$, It takes two clone samples $(C_i, C_j)$, corresponding AST sub-trees $(S_i,S_j)$ and feature vectors $(V_i, V_j)$ representing two code clones as inputs (line 1-4 in Algorithm~\ref{alg1}), and we utilize a helper function $LCS()$ to find the Longest Common Subsequence between two lists of sub-trees.

When the code clone samples are symbolically executed, we start by checking if the constraints, obtained from previous formal verification step, are equivalent. Then the feedback procedure after is conducted as two folds: 

(1) If they indeed share the same constraints, we remove the uncommon subtrees (where can be treated as numerical weight as 0) as we now know they will not affect the output of constraints (line 10-13). This process is to make sure the remaining trees are identical so that they will be detected as code clone in the future.

(2) If they have different constraints, we obtain the uncommon subtrees from $(S_i, S_j)$(line 15-17) and add numerical weight, $\delta > 1.0$, one by one. We iterate the list and we trace back to the vector using the vector index to adjust the weight $\delta$ for that specific location correspondingly (line 18-22). We initialize the weight $\delta$ as a random number which is greater than 1.0 and re-calculate the euclidean distance between two feature vectors. We repeat this process until the distance is out of current code similarity threshold $S$ (line 19-20). This is designed to guarantee that these two code samples will not be considered as code clone in the future.
Finally, the feedback can run in a loop fashion to eliminate false positives. The termination condition for our feedback loop is that no more false positives can be further eliminated or observed.

{\it Example:} Here, we give an example to illustrate how our formal feedback works. We use the false positive example showing in Figure~\ref{as_fp}. 
Assuming the feature vectors are $<7,2,2,2,0,1,1,1,1>$ and $<8,1,1,2,1,1,1,1,1>$ respectively, where the ordered dimensions of vectors are
occurrence counts of the relevant nodes: \textbf{ID, Constant, ArrayRef, Assignment, StrucRef, BinaryOp, UnaryOp, Compound,} and \textbf{For}. Based on the threshold defined in equation~\ref{clones}, these two code fragments will be clustered as clones when $S = 0.75$.
During the feedback loop, we first identify these 2 different nodes in each tree by finding the LCS. Assuming we initial the weight $\delta = 2$ and add it to the corresponding dimension in the feature vectors, we can obtain the updated feature vectors as $<7,1+1\times\delta,1+1\times\delta,1+1\times\delta,0,1,1,1,1>$ and $<7+1\times\delta,1,1,2,1\times\delta,1,1,1,1>$. We then re-calculate the euclidean distance of these two updated feature vectors, and they will be no longer satisfied within the threshold $\sqrt{2 (1-S) \times min (S(C_i), S(C_j))}$. Thus, we can eliminate such false positives in the future.


It is also worth mentioning that our feedback algorithm has enabled a closed-loop learning-based operation to improve the scalability of our pointer-related code clone detection framework. Because this method adds benefits from formal analysis and can significantly reduce the false positives without human efforts involved. Here, we use pointer analysis as an example to explain our framework. In addition, our feedback algorithm can be adjusted to different domains with user-defined policies.
	\section{Implementation}
\label{Implementation}
This section discusses our implementation of \NAME\ and how we integrate the tools we used.

\textbf{Program Slicing:} We instrument a static code analysis tool, Joern~\cite{joern}, for our program slicing module.
Joern is able to store code property graphs (like ASTs) in a Neo4J graph database~\cite{webber2012programmatic}, here we call it AST database, for user to write their own scripts to do static code analysis. We develop a python script to build ASTs for each function and construct dependency graphs. After, we store them into Neo4J graph database for further analysis.
As Joern cannot store the source code location information, such as which lines these statements is from in the source code. We instrument Joern to include additional information for a certain statement using a C++ script, including file path along with code line number, so that we can trace back to source code after we perform static program slicing to isolate original source code into pointer isolated functions.

\textbf{Code Clone Detection:} DECKARD~\cite{jiang2007deckard}, a static Code Clone Detection tool, is used for code clone detection in \NAME. DECKARD is a tree-based code clones detection tool that computes certain characteristic vectors within code parse trees and then clustering these vectors depending on their Euclidean distances. We instrumented DECKARD interfaced with our program slicing module to automate the clone detection process.

\textbf{Clone Verification:} We instrument a source code symbolic execution tool, KLEE~\cite{cadar2008klee} and SMT solver Z3~\cite{zheng2013z3} for our clone verification module. We first develop a python script to automatically add codes into the pointer isolated code fragments and make pointer-related variables symbolic using KLEE provide library function. We then deploy the symbolic executor in KLEE for a target location to start performing symbolic execution in the source code, beginning with the starting line of code and execute till the ending line of code in the code fragment.
Finally, we implemented our feedback in also python based on the algorithm proposed in Section~\ref{feedback}.

	\section{Evaluation}
\label{eva}
In this section, we evaluate the effectiveness of our approach. First, we describe a detailed evaluation results of \NAME\ against a tree-based code clone detection tool DECKARD~\cite{jiang2007deckard} in terms of code clone detection. Then, we conduct various case studies for applications security analysis. Finally, we analyze unreported bugs in Links version and LibreOffice.

\subsection{Experiment Setup}
We implemented our approach in a tool, called \NAME. The first segment is to select 7 different benchmarks from real-world applications: bzip2, hmmer and sphinx3 from SPEC2006 benchmark suite~\cite{spec}; man and gzip from Bugbench~\cite{lu2005bugbench}; thttpd-2.23beat1~\cite{thttpd_ACME}, a well-known lightweight sever and a lightweight browser links-2.14~\cite{links}. All the experiments are conducted on a 2.54 GHz Intel Xeon(R) CPU E5540 8-core server with 12 GByte of main memory. The operating system is ubuntu 14.04 LTS.

In the second segment, to
configure DECKARD, we used the parameter settings projected by Jiang
et al. ~\cite{jiang2007deckard}, setting minimum token number (minT) as 20, stride to infinite, and code similarity is set between 0.70 and 1.0.

\subsection{Code Clones Detection }
\begin{table}[H]
	\centering

\footnotesize
	\begin{tabular}{|c|c|c|c|c|}
		\hline
		\textbf{Bench.} & \textbf{Program Size } & \textbf{\#Code clones} & \textbf{\#Code clones} &\textbf{\% Code clones} \\ 
		& \textbf{(LoC)} & \textbf{without slicing} & \textbf{{Our approach}} &\\ \hline
		\textit{bzip2}     & 5,904                       & 432                            & 1,084         &        150.92\%         \\
		\textit{sphinx3}   & 13,207                      & 1,047                            & 3,546   &         238.68\%          \\
		\textit{hmmer}     & 20,721                      & 1,238                         & 4,391      &          254.68\%           \\
		\textit{thttpd}    & 7,956                       & 611                           & 1,398     &             128.80\%      \\
		\textit{gzip}      & 5,225                       & 36                              & 365        &            913.89\%       \\
		\textit{man}       & 3,028                       & 47                              & 443        &          842.55\%          \\
		\textit{links}     & 178,441                     & 3,007                            & 9,809     &            226.21\%         \\ \hline
	\end{tabular}
\vspace{0.1in}
\caption{Comparison of number of code clones detected before and after using our approach }
\label{statistics}
\end{table}
\begin{table}[H]
	\centering
\tiny
\begin{tabular}{|c|c|c|c|c|c|}
	\hline
	\multirow{2}{*}{\textbf{Bench.}} & \multirow{2}{*}{\textbf{\begin{tabular}[c]{@{}c@{}}Pointer related \\ LoC\end{tabular}}} & \multicolumn{2}{c|}{\textbf{Clone Detection w/  DECKARD}} & \multicolumn{2}{c|}{\textbf{Clone Detection w/ Our Approach}} \\ [2ex]\cline{3-6} 
	&                                                                                              & \textbf{\# Cloned LoC}      & \textbf{\% Cloned LoC}      & \textbf{\# D.S LoC}           & \textbf{\% D.S LoC}           \\ \hline
	\textit{bzip2}                      & 3,279                                                                                         & 1,066                        & 32.51\%                     & 2,038                          & 62.15\%                       \\ [2ex]\hline
	\textit{sphinx3}                    & 9,519                                                                                         & 3,073                        & 32.28\%                     & 7,224                          & 75.89\%                       \\ [2ex]\hline
	\textit{hmmer}                      & 11,635                                                                                        & 3,163                        & 27.19\%                     & 6,929                          & 59.55\%                       \\ [2ex]\hline
	\textit{thttpd}                     & 4,390                                                                                         & 1,279                        & 29.13\%                     & 2,267                          & 51.64\%                       \\ [2ex]\hline
	\textit{gzip}                       & 2,289                                                                                         & 219                         & 9.57\%                      & 919                           & 40.15\%                       \\[2ex] \hline
	man                                 & 1,683                                                                                         & 248                         & 14.74\%                     & 826                           & 49.08\%                       \\ [2ex]\hline
	links                               & 28,334                                                                                        & 6,429                        & 22.69\%                     & 18,334                         & 64.71\%                       \\ [2ex]\hline
\end{tabular}
	\caption{Comparison of code clone coverage between DECKARD and our approach}
	\label{cloned_loc}
\end{table}
In the third segment, we estimate code clone quantity using number of code clones that are detected before and after we use \NAME\ for pointer analysis purpose. We evaluate the experiments in the following ways: code clones quantity, the flexibility of code similarity configuration and false positives analysis.

\begin{figure*}
	\centering

	\subfloat[][thttpd]{\includegraphics[scale=0.2]{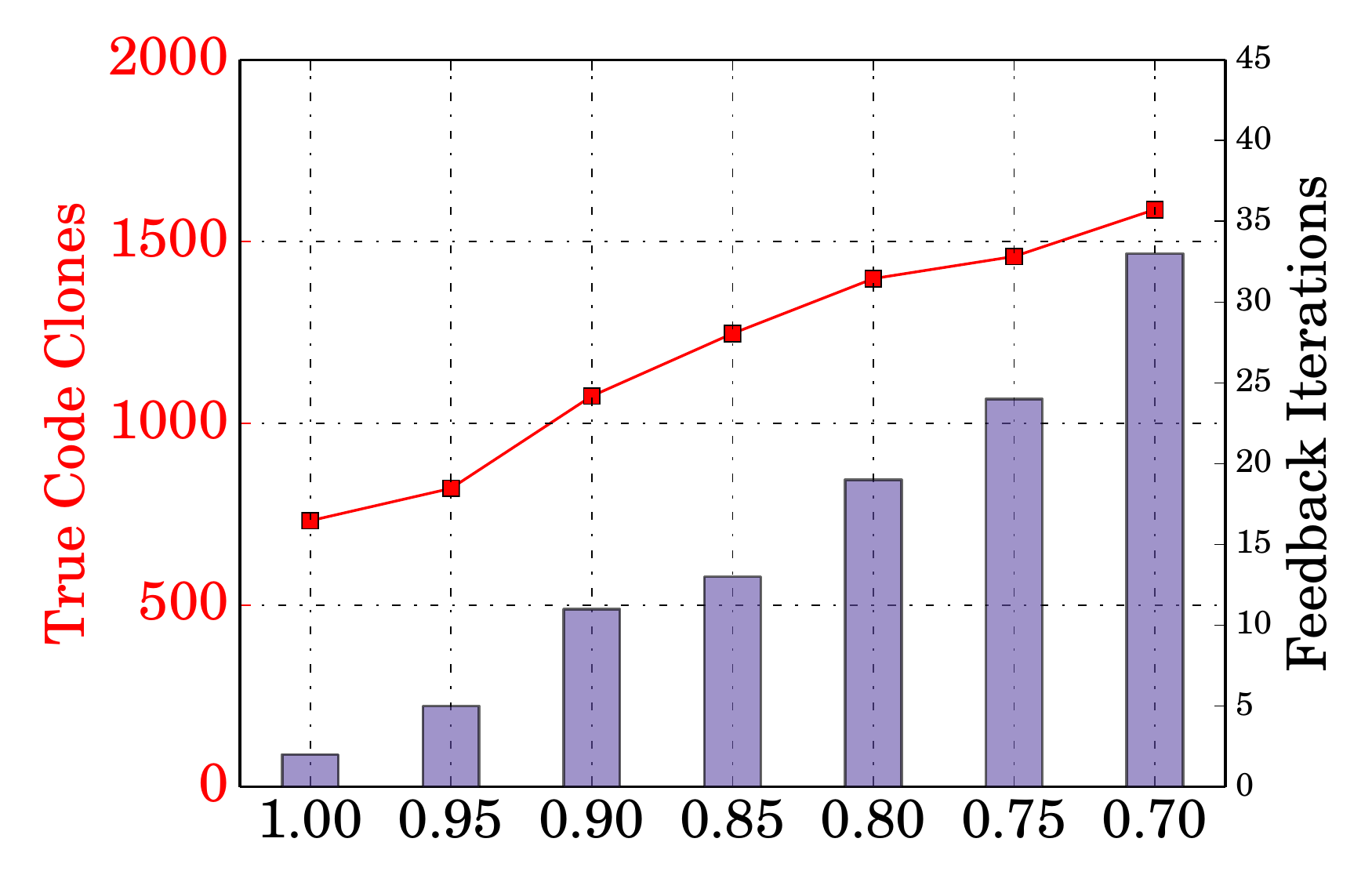}\label{<figure6>}}
	\hspace{-0.1in}
	\subfloat[][links]{\includegraphics[scale=0.2]{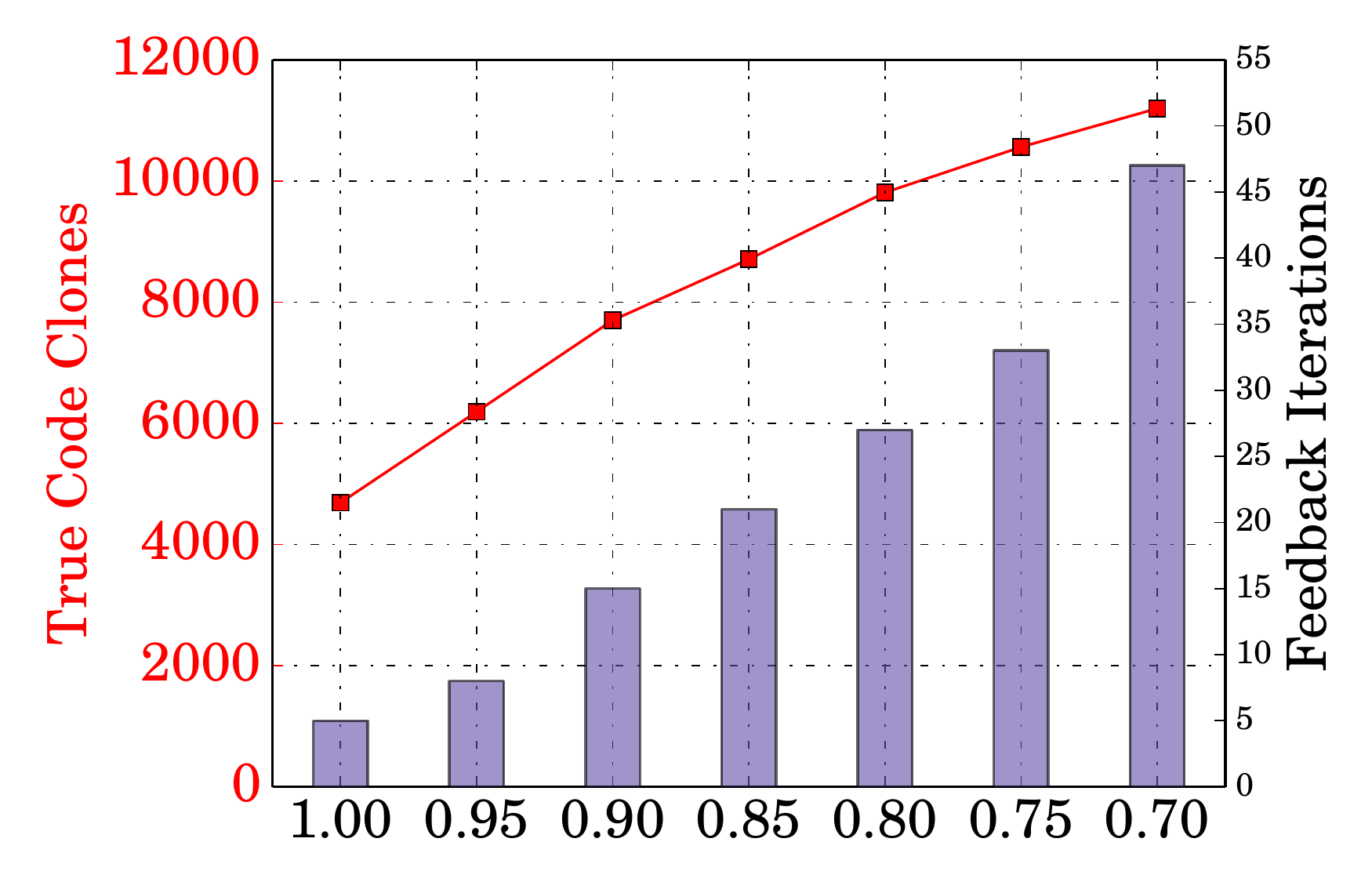}\label{<figure7>}}
	\subfloat[][bzip2]{\includegraphics[scale=0.2]{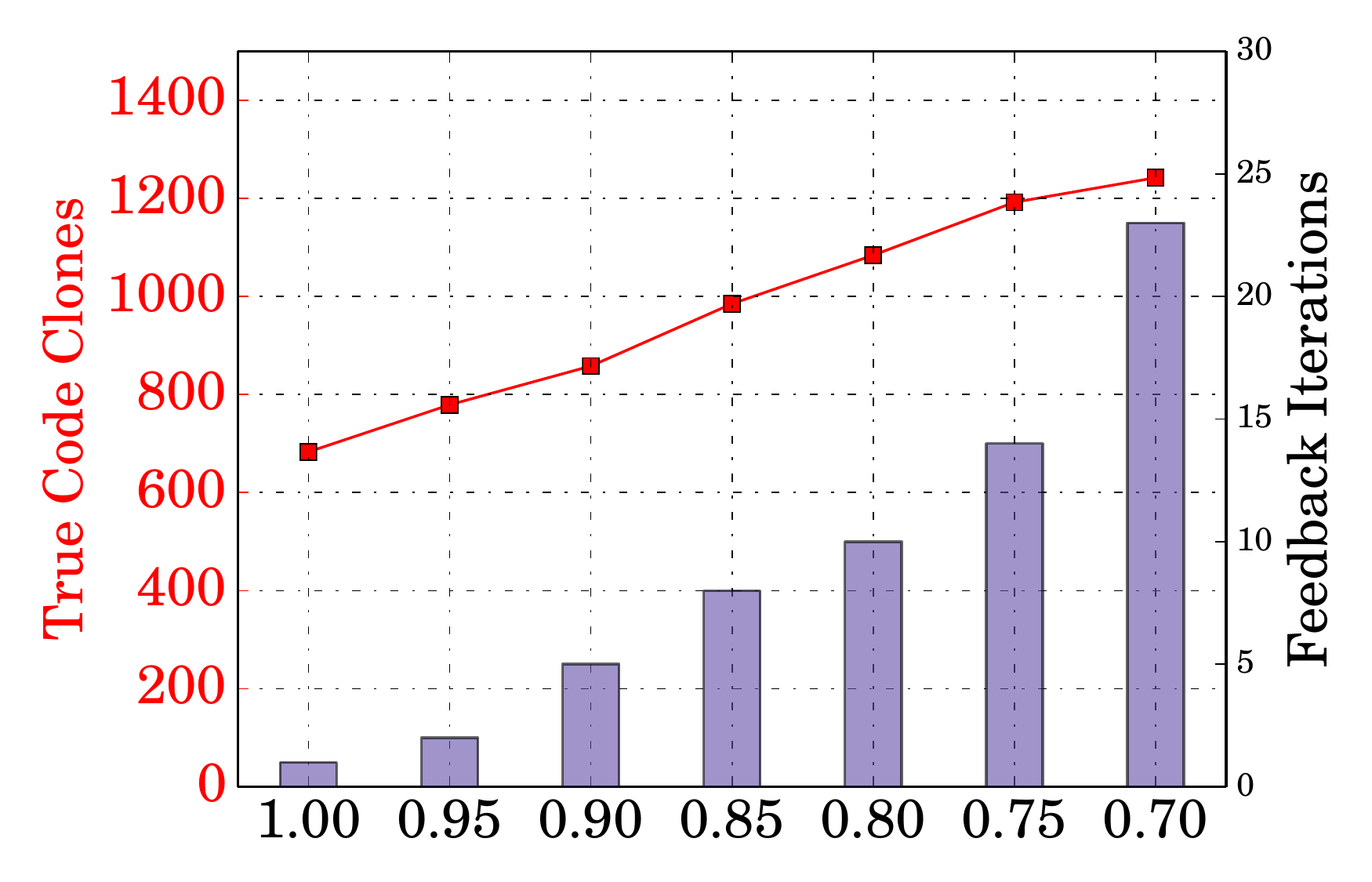}\label{<figure8>}}
	\subfloat[][sphinx3]{\includegraphics[scale=0.2]{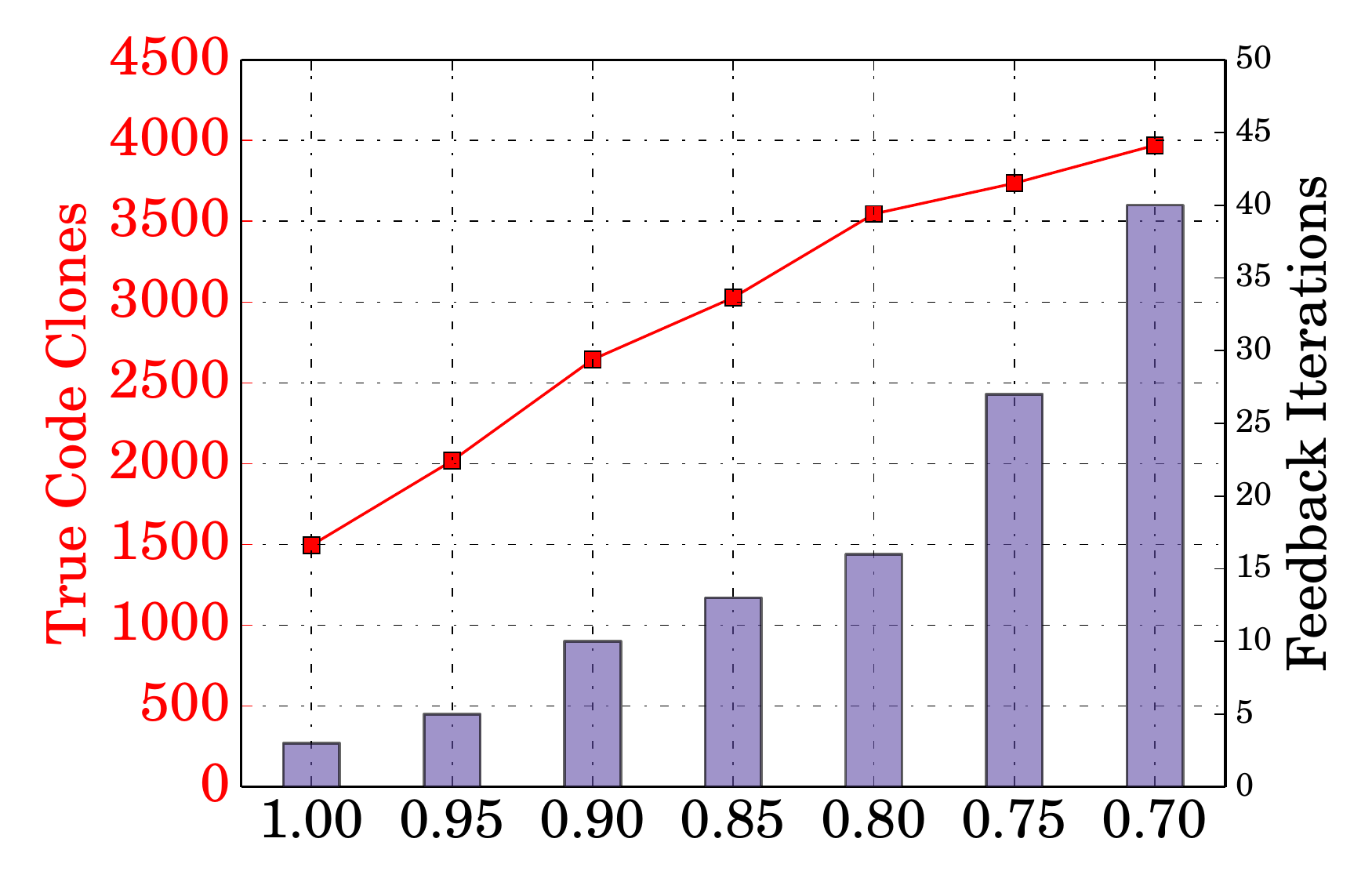}\label{<figure9>}}
				\hspace{-0.1in}
	\subfloat[][hmmer]{\includegraphics[scale=0.2]{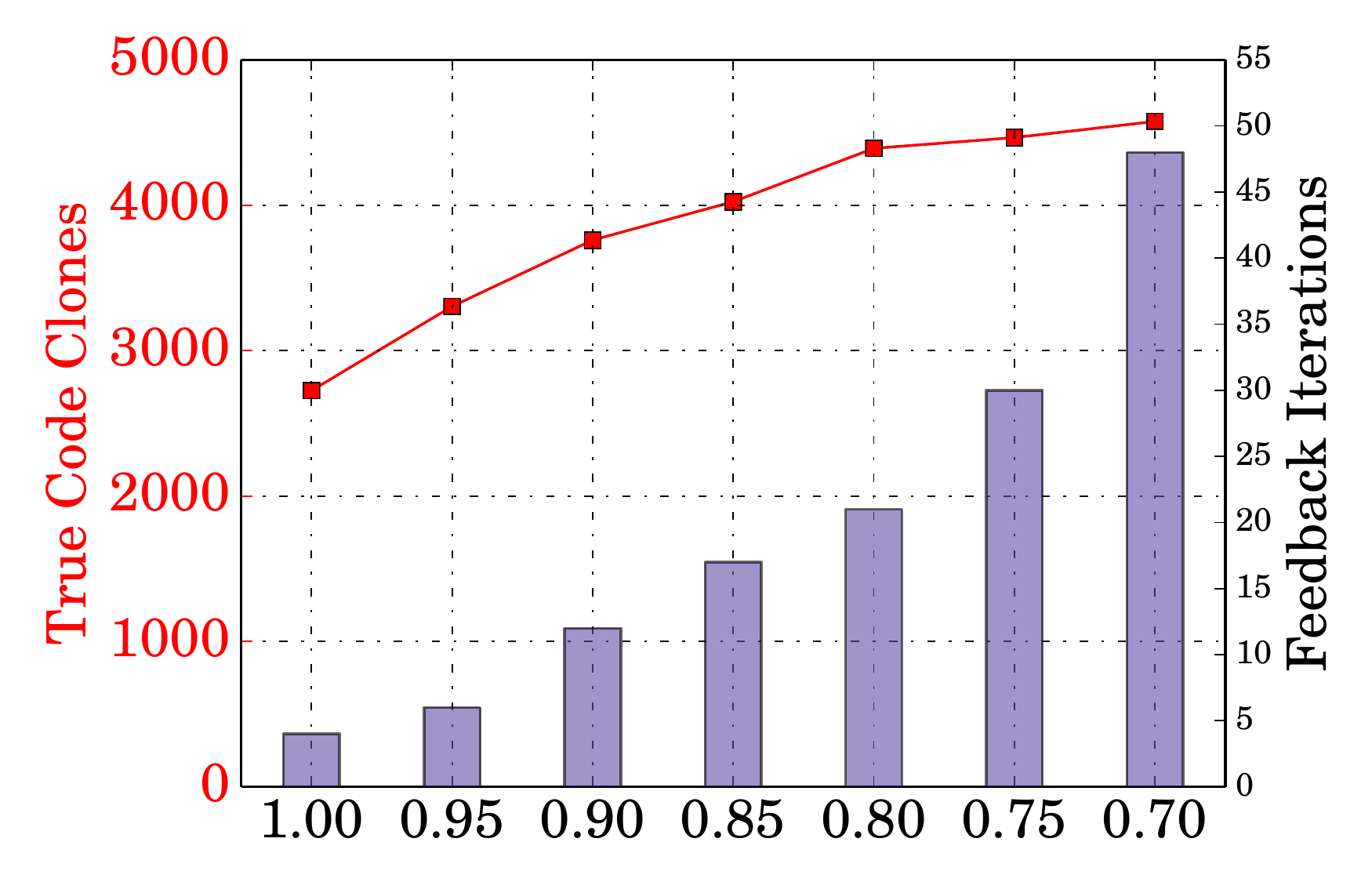}\label{<figure10>}}
	\subfloat[][man]{\includegraphics[scale=0.2]{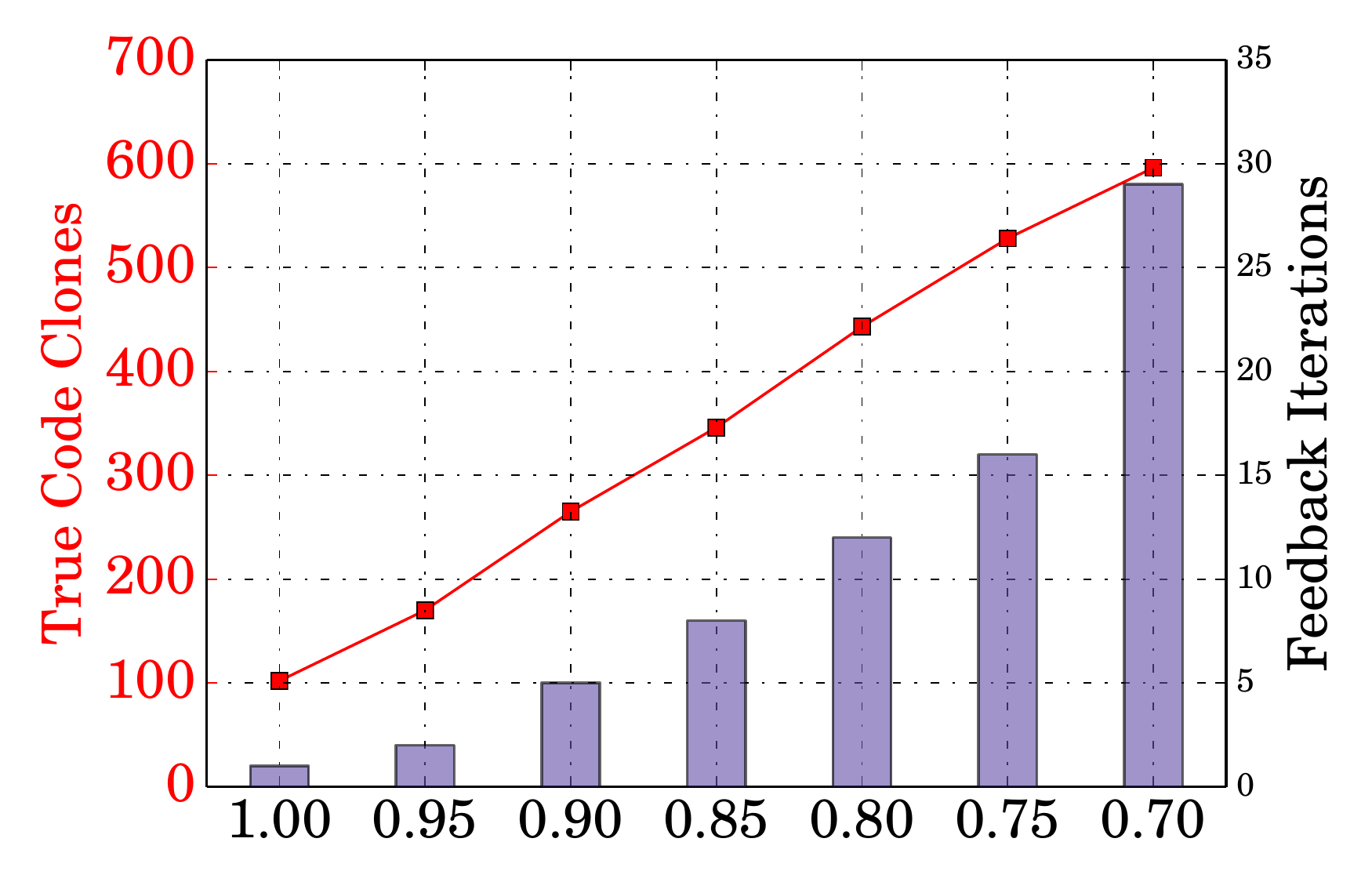}\label{<figure11>}}
	\subfloat[][gzip]{\includegraphics[scale=0.2]{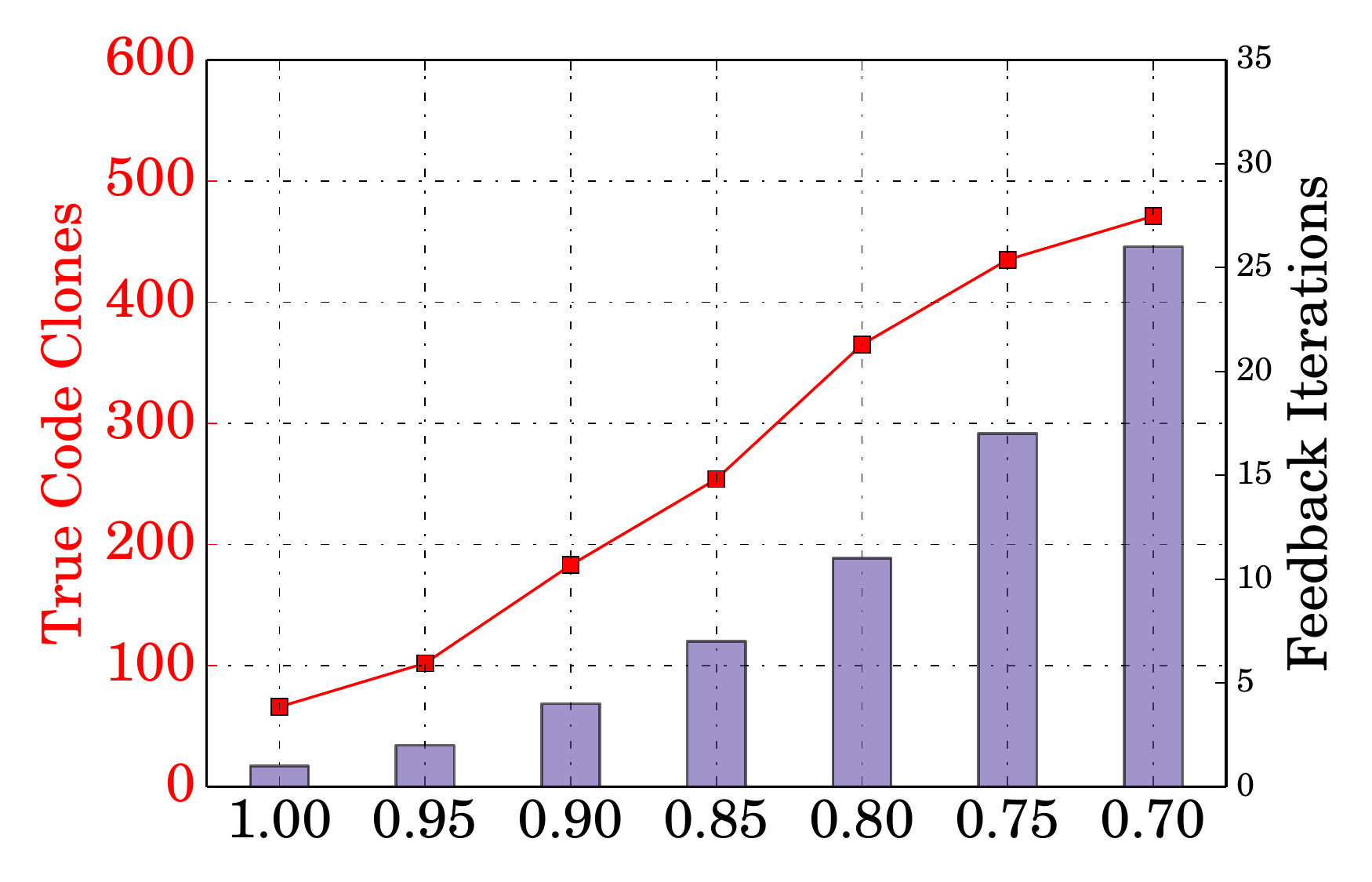}\label{<figure12>}}
	\caption{The amount of code clones detected from \NAME\ with the number of iterations for feedback until converge after relaxing the code similarity from 0.70 to 1.00}
	\label{CodeClones}
\end{figure*}
%
%

We assessed the effectiveness of \NAME\ to illustrate the optimal results \NAME\ can achieve. The code similarity is set as 0.80 with feedback enabled to eliminate false positives until converge (no more false positives can be observed or eliminated) in the first experiment. Table~\ref{statistics} determines the size of the corresponding percentage of more code clones detected using our approach. As can be seen from the results that \NAME\ is able to detect 393.68\% more code clones in average compared to the clone detection without slicing and feedback, with the lowest as 128.80\% in \textit{thttpd} and highest up to 913.89\% in \textit{gzip}. Note that our approach achieves the best performance in two smaller benchmark \textit{gzip} and \textit{man}. The reason being number of identical code clones is relatively small in both applications (36 in \textit{gzip} and 47 in \textit{man} respectively). While using our approach, we harness the power of program slicing and feedback using formal analysis, which permits us to detect more true code clones.

In the fourth segment, we add an additional experiment to address the clone coverage. The objective for clone coverage is to estimate the fraction of a program that is detected cloned code with our optimal configuration. In this case, we only evaluated the coverage of code clones detected in terms of pointer-related code. We estimated the total number of pointer-related code lines cross the entire program and the detected clone lines using DECKARD and our approach can be seen in Table~\ref{cloned_loc}. It presents the total detected
pointer related cloned lines, named as \textit{Domain Specific LoC} (D.S LoC), using our approach. The percentage of D.S LoC ranges from 40.15\% to 75.89\%, while for DECKARD the number ranges from 9.57\% to 32.51\%. The results indicate the difficulty in comparing the coverage for different applications as the results are usually sensitive to: (1) the type of application, such as sphinx3 has intensive pointer access, thus it has the highest clone coverage using our approach; (2) the different configurations may lead to different results, since here we set up code similarity as 0.80. However, this experimentation is to show that there is a considerable amount of code clones in large code bases in general and our approach can effectively detect such clones and outperform previous approaches.

In the fifth segment, we relaxed the code similarity threshold from 0.70 to 1.00 to show our approach is capable to detect many more code clones within a flexible user-defined configuration. However, it is reasonable to expect more false positives to occur while we are using smaller code similarity. Additionally, we implemented our code clone detection based on DECKARD, which is a syntax tree-based tool and may report semantically
different but syntactically similar code as clones causing more false positives. Note that false negatives
occur if two clone samples have different constraints but are actually the same expression after being solved by the constraint solver. However,
false negatives only result
in actually true clones being missed by \NAME\ and are not critical in security perspective. Thus, we do not evaluate \NAME\ for
false negatives in our study.

We enable a closed-loop feedback to vector embedding in tackling such false positives issue as mentioned in the preceding section. So, we analyzed  the effectiveness of our feedback mechanism in terms of eliminating the false positives. 
In this experiment, we applied our feedback as soon as we observed two code clone samples having different constraints obtained from symbolic execution through our clone verification process. We executed several iterations of our feedback until the percentage of false positives converged (no more false positives can be eliminated or observed). Figure~\ref{CodeClones} shows the number of true code clones detected across all the benchmarks from our approach (drawn as red line in each figure) and the number of iterations for feedback needed to converge (shown as the bar plot in each figure) correspondingly. In the next step, we replicated the same experiments with three different code similarities setups in other smaller benchmarks. 
As expected, it takes more iterations for the feedback to converge with smaller code similarity among all benchmarks, and we are still able to detect more true code clones while we reduce the code similarity. However, the results show there is no substantial improvement in terms of the number of true code clones increased after code similarity is set as smaller than 0.80. As mentioned in previous section, the code similarity is defined as  $S (T_1, T_2) = \frac{2S}{2S+L+R}$, where $S$ is the number of shared AST nodes in $T_1$ and $T_2$, $L$ and $R$ are the different nodes in two code clone samples. At least 20\% of the AST nodes are different while the code similarity equal to 0.80.

\subsection{Feedback for False Positives Elimination}
\begin{figure}[ht]
	
\centering 
\includegraphics[scale=0.5]{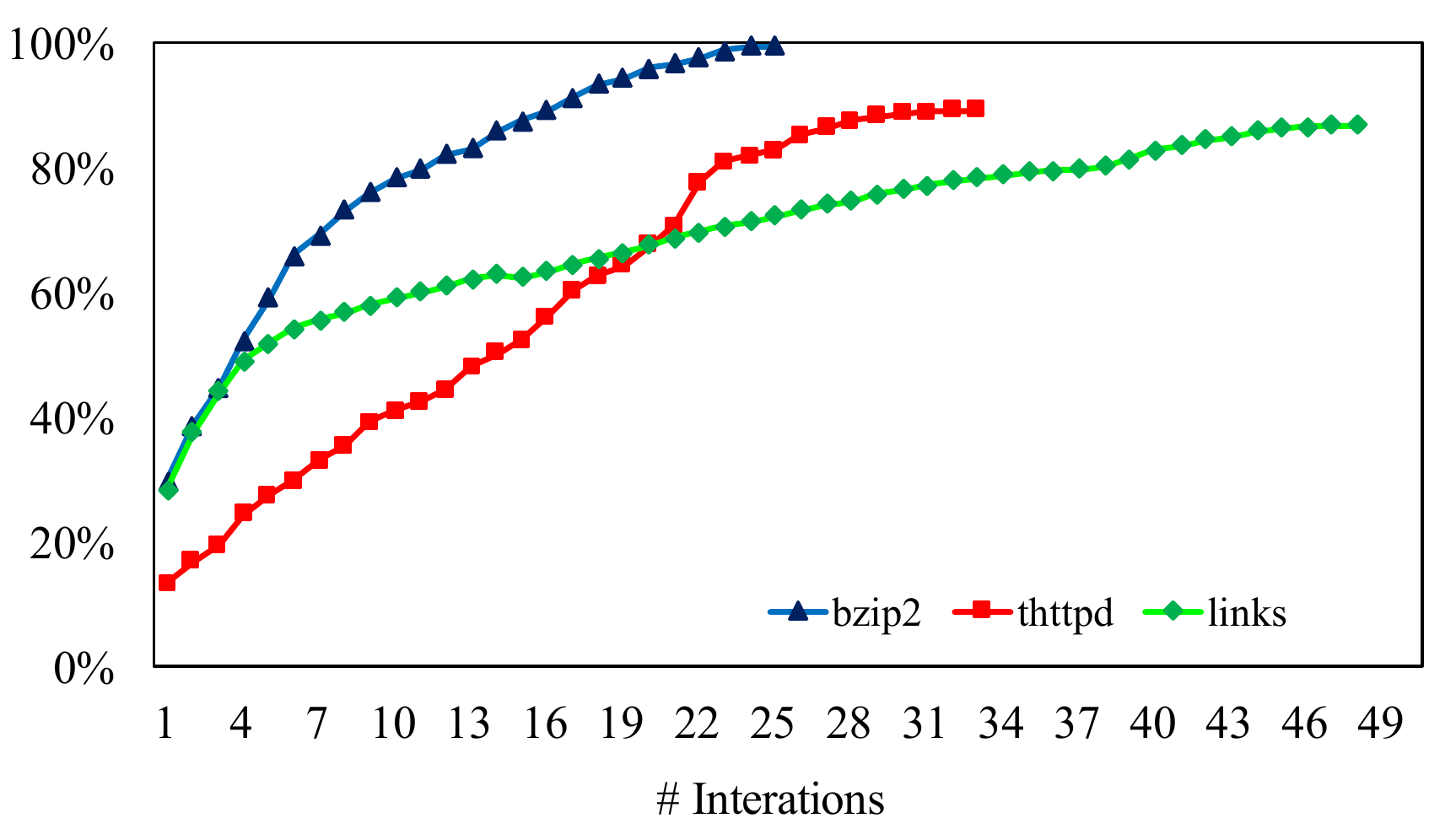}

\caption{Accumulated percentage of false positives eliminated by \NAME\ with code similarity set to 0.70}
\label{fig:fp_removed}
\end{figure}
We analyzed the number of false positives that could be eliminated by our approach. Here, we chose bzip, thttpd and Links as representative applications to show the results. Figure~\ref{fig:fp_removed} presents the accumulated percentage of false positives eliminated by \NAME\ in each iteration with Code Similarity set to 0.7. Here, we are able to eliminate 99.32\%, 89.0\%, and 86.74\% of false positives in bzip2, thttpd and Links respectively.

The results show our feedback mechanism can effectively remove the majority of false positives admitted from code clone detection. The performance of our feedback is sensitive to different programs due to different program behaviors and program size. As the results show, more feedback iterations are needed for larger program in general (e.g. 26 iterations for bzip2 to converge while 48 iterations for Links, as Links is much larger than bzip2). On the other hand, the number of iterations can also be affected by our clone verification module since we use random sampling approach. Based on the experiment results, we cannot normalize a common removal ratio pattern cross different programs. For instance, 29.83\% of false positives can be eliminated at the first iteration for bzip2, the number is only 13.35\% for thttpd instead. Finally, our feedback may not be able to remove 100\% of false positives, that is because there are several special cases that we cannot remove them using current implementation, such as multiple branches or indirect memory access with the value of array index derived from another pointer. 
\subsection{Bug findings }
\begin{table*}[h]
	\centering
	\tiny
	\begin{tabular}{c|c|c|c|c}

		\hline
		\textbf{Bug Type} & \textbf{Source File}                                                                               & \textbf{Function Name}    & \textbf{Pointer Name}    & \textbf{Bug Report}                                                                                                                                                                                                                           \\ [2ex]\hline

		Null Dereference  & Links-1.4/language.c                                                                               & get\_language\_from\_lang & lang                     & Not Reported           
		 \\ [2ex]\hline

		Null Dereference  & Links-1.4/language.c                                                                               & get\_language\_from\_lang & p                        & Not Reported                                                                                                                                                                                                                                                       \\ [2ex]\hline
		Null Dereference  & Links-1.4/connect.c                                                                                & make\_connection          & host                     & Not Reported            
		\\ [2ex]\hline
		Memory Leak       & Links-1.4/ftp.c                                                                                    & ftp\_logged               & rb                       & Not Reported              
		\\[2ex] \hline
		Memory Leak       & Links-1.4/bfu.c                                                                                    & do\_tab\_compl            & items-\textgreater{}text &Silently patched                                                                                                                                                                                                                                         \\[2ex] \hline
		Memory Leak       & Links-1.4/terminal.c                                                                               & add\_empty\_window        & ewd                      & Not Reported                                                                                                                                                                                                                                                              \\[2ex] \hline
		Buffer Overflow   & \begin{tabular}[c]{@{}l@{}}libreoffice-6.0.0.1/sw/\\source/filter/ww8\\/ww8toolbar.cxx\end{tabular} & SwCTBWrapper::Read        & rCustomizations          & Publicly patched                                                                                                                                                                                        \\[2ex] \hline
	\end{tabular}
	\vspace{0.1in}
	\caption{Using our approach to test Links-1.4 and libreoffice-6.0.0.1}
	\label{bugs}
\end{table*}
One benefit of our approach is to use a clone-based approach to enable a rapid security analysis. In this experiment, we use \NAME\ to detect potential vulnerabilities existing in the applications. We use Links version 2.14 and LibreOffice version 6.0.0.1 as representative benchmarks. In particular, we
discovered 6 unique and unreported bugs in Links, including 3 memory leaks and 3 null dereference vulnerabilities. five out of six of the bugs have not been found before, and one of the memory leaks bug has been silently patched in the newer version of Links.

Table~\ref{bugs} shows the
details of these bugs found by our method. Here we show three types of bug examples, null dereference bugs, memory leak and buffer overflow. 
\subsubsection{Links Case Study}

\begin{figure}[h]
\hspace{0.3in}
	\begin{minipage}[b]{0.92\linewidth}
		\begin{lstlisting}[style=CStyle, frame=single,,escapechar=\%]
int get_language_from_lang(unsigned char *lang) {
unsigned char *p;
int i;
%\Hilight%lang = stracpy(lang);
%\Hilight%//Uncheck the memory allocation
%\Hilight%lang[strcspn(cast_const_char lang, ".@")] = 0;
if (!casestrcmp(lang, cast_uchar "nn_NO"))
	strcpy(cast_char lang, "no");
...
search_again;
for (i = 0; i < n_languages(); i++) {
	p = cast_uchar translations[i].t[T__ACCEPT_LANGUAGE].name;
	if (!p)
		continue;
%\Hilight%	p = stracpy(p);
%\Hilight%//Uncheck the memory allocation
%\Hilight%	p[strcspn(cast_const_char p, ",;")] = 0;
	if (!casestrcmp(lang, p)) {
		mem_free(p);
		mem_free(lang);
		return i;
	}
	mem_free(p);
}
...
mem_free(lang);
return -1;
}
\end{lstlisting}
		\caption{Null Dereference bugs in function \textit{get\_language\_from\_lang} of the lightweight browser Links}
				\label{null_dereference}
	\end{minipage}		
\end{figure}
In the first case study, we employ \NAME\ to uncover vulnerabilities in Links, a lightweight browser. The results show \NAME\ finds 6 unreported bugs in Links version 2.14, including 3 memory leaks and 3 Null dereference vulnerabilities. And 1 of the memory leaks bug is silently patched in the newer version of Links. 

As an example, let us consider the function \textit{get\_language\_from\_lang} shown in Figure~\ref{null_dereference}. This function is implemented as setting language from local serves. This function provides an illustrative example because the programmer confirms that the $stracpy$ requires validation in the comment on
line 5
The $stracpy$ function in line 4 is implemented as dynamic memory allocation for a pointer. The bug arises when the code fails to allocate memory to pointer $lang$ using $stracpy$ function and return $NULL$ to pointer $lang$. Thus, there is a potential null pointer dereference in line 6.

After we deploy program slicing and code clone detection, we are also able to identify the same bug with the assistance of symbolic execution rapidly for pointer $p$ in line 15, as two code snippets are identified code clones (line 4-6 and line 15-17).
Similarly, pointer $p$ is unchecked after memory allocation, which results in the same vulnerability existing in the codes. This example shows the advantage of our approach combining program slicing and code clone detection for vulnerability discovery.

%
%
%
%
%
\subsubsection{LibreOffice Case Study}

\begin{figure}[h]
	\hspace{0.3in}
	\begin{minipage}[b]{0.9\linewidth}
		\begin{lstlisting}[style=CStyle, frame=single,,escapechar=\%]
bool SwCTBWrapper::Read( SvStream& rS )
{   
...
if (cCust)
{   
	...
	for (sal_uInt16 index = 0; index < cCust; ++index)
	{   
		Customization aCust( this );
		if ( !aCust.Read( rS ) )
			return false;
		rCustomizations.push_back( aCust );
	}
}
...
std::vector< sal_Int16 >::iterator it_end = dropDownMenuIndices.end();
%\Hilight%for ( std::vector< sal_Int16 >::iterator it = 
%\Hilight%dropDownMenuIndices.begin(); it != it_end; ++it )
{   
%\Hilight%	rCustomizations[ *it ].bIsDroppedMenuTB = true;
}
return rS.good();
}
\end{lstlisting}
\caption{Source Code from function \textit{Links::SwCTBWrapper::Read} where a buffer overflow bug via pointer $rCustomizations$}
\label{fig:bug_3}
	\end{minipage}	
\end{figure}
In the second case study, LibreOffice is an open source office tool, which is written in multiple programming languages including C/C++ and Java. Currently, our approach is working to C/C++ code only. Thus, we only deployed our approach on the C/C++ files in LibreOffice. Our approach was able to identify a heap-based buffer overflow bug in function \textit{Links::SwCTBWrapper::Read}. Figure~\ref{fig:bug_3} shows the original source code. \NAME\ identified a group of code clones of code snippets from line 17-21 in the same cluster. Our feedback mechanism eliminated the other code clones as false positives after 16 iterations.

This function is used to read a crafted document containing a Microsoft Word record (named as a structural array $rCustomizations$ in the source code) from beginning to the end. The size of structural array $rCustomizations$ is defined as $static\_cast<sal\_Int16>( rCustomizations.size() )$. However, the \textit{for} loop in line 17, it does not do a propel bound check of a customizations array index. The value of $*it$ could be negative or larger than the size of  $rCustomizations$. When our approach deploys partial symbolic execution for this $for$ loop, it will yield potential buffer overflow error \footnote{However, after we started our research, this bug has been found earlier of 2018 and public patched in the newer version of LibreOffice. More details about this bug can be found in the report CVE-2018-10120~\cite{libreoffice}}.

	\section{Related Work}
\label{related}
Related works including code clone detection and program slicing have been discussed closely throughout the paper. In this section, we summarize some additional related work. We focus on
existing static code analysis and code clone detection approaches. Other approaches for vulnerability discovery will be also discussed in this section.

\textbf{Code clone detection.} Different approaches for code clone detection have been proposed. Recall that detection techniques generally can be classified into several categories. First, text-based or simple string matching based techniques~\cite{ducasse1999language,baker1995finding,baker1997parameterized} apply slight program transformations and apply a single code similarity measurement by comparing sequences of text. Such text-based techniques are limited in the scalability in large code bases and only finding exact match code clone pairs. Second, tree or token-based clone detections~\cite{kontogiannis1996pattern,wahler2004clone,baxter2004dms,zhang2014personal} are proposed by parsing program into tokens or generate abstract syntax trees representation of the source
program. Consequently, tree or token-based approaches usually more robust against code changes. Some well-known tools like CC-Finder~\cite{kamiya2002ccfinder}, DECKARD~\cite{jiang2007deckard} and CP-Miner~\cite{li2006cp}. 
However, above approaches are still not sufficient to non-contiguous and intertwined code clone. 


\textbf{Pointer analysis and symbolic execution.} In this paper, we choose pointer analysis as our analysis domain. Invalid use of pointers can lead to hard-to-find bugs and may expose security vulnerabilities. Thus, analyzing them is critical for software analysis as well as optimization. Conventional pointer analysis executes in an exhaustive way in order to analyze every pointer in the code. Previous works on pointer analysis~\cite{emami1994context,heintze2001demand,nystrom2004bottom} has indicated that the main bottleneck towards scalability for exhaustive pointer analysis. For instance, dynamic runtime bound checking performs exhaustive pointer analysis to detect out-of-bound array accesses. In the case of memory safety and pointer risky usage, symbolic execution can overcome these by bound checking all the pointer dereference, actively exploring all the possible paths and conditions in the code~\cite{avgerinos2014automatic,godefroid2012sage}. Nevertheless, symbolic execution has limited scalability and becomes extremely time consuming due to states/paths explosion, especially in larger programs.

\textbf{Intergrated learning framework for bug findings.}  Statistical method and formal method combined framework have been stuided ~\cite{hu2017binary,xue2020learn2reason,pewny2015cross,xue2018clone,xue2020twin,xue2018clone_hunter,xue2018morph,xue2019hecate,chen2020chop}. 
StatSym~\cite{yao2017statsym} and SARRE~\cite{li2016sarre} propose frameworks combining statistical and formal analysis for vulnerable path discovery. SIMBER~\cite{xue2017simber} proposes a
statistical inference framework to eliminate redundant bound
checks and improve the performance of applications without
sacrificing security. Another line of work use Natural Language Processing and machine learning to bug detection. For example, Chucky~\cite{yamaguchi2013chucky} uses context-based
Natural Language Processing to detect missing check vulnerability. These techniques, often transfer code into intermediate representation and then rely on static code analysis to find bugs. In this paper, we develop an integrated framework that harness the effectiveness of code clone detection and formal analysis techniques for a rapid security analysis on source code at scale. In contrast to pure formal analysis, such as symbolic execution, we are able to achieve a significant speedup to find vulnerabilities.

	\section{Conclusion}
\label{conclusion}

In this paper, we porpose an integrated reasonding engine, \NAME, for pointer-related code clone detection in source code. \NAME\ can automatically identify pointer-related codes from large code bases and perform code clone detection to enable a rapid security analysis. The evaluation results show \NAME\ can detect up to 9$\times$ more code clones comparing to conventional code clone detection approaches and can remove 91.69\% false positives in average. We further conduct security case studies for memory safety issues. In particular, we show that using \NAME, we are able to find 6 unreported bugs in an open source web browser Links version 2.14 and one public patched bug in an open source office tool libreOffice-6.0.0.1. 
\bibliographystyle{elsarticle-num}

	\bibliography{REF}

\begin{thebibliography}{10}
\expandafter\ifx\csname url\endcsname\relax
  \def\url#1{\texttt{#1}}\fi
\expandafter\ifx\csname urlprefix\endcsname\relax\def\urlprefix{URL }\fi
\expandafter\ifx\csname href\endcsname\relax
  \def\href#1#2{#2} \def\path#1{#1}\fi

\bibitem{gabel2010study}
M.~Gabel, Z.~Su, A study of the uniqueness of source code, in: Proceedings of
  the eighteenth ACM SIGSOFT international symposium on Foundations of software
  engineering, ACM, 2010, pp. 147--156.

\bibitem{kim2005empirical}
M.~Kim, V.~Sazawal, D.~Notkin, G.~Murphy, An empirical study of code clone
  genealogies, in: ACM SIGSOFT Software Engineering Notes, Vol.~30, ACM, 2005,
  pp. 187--196.

\bibitem{li2006cp}
Z.~Li, S.~Lu, S.~Myagmar, Y.~Zhou, Cp-miner: Finding copy-paste and related
  bugs in large-scale software code, IEEE Transactions on software Engineering
  32~(3) (2006) 176--192.

\bibitem{xue2019machine}
H.~Xue, S.~Sun, G.~Venkataramani, T.~Lan, Machine learning-based analysis of
  program binaries: A comprehensive study, IEEE Access 7 (2019) 65889--65912.

\bibitem{baker1997parameterized}
B.~S. Baker, Parameterized duplication in strings: Algorithms and an
  application to software maintenance, SIAM Journal on Computing 26~(5) (1997)
  1343--1362.

\bibitem{kamiya2002ccfinder}
T.~Kamiya, S.~Kusumoto, K.~Inoue, Ccfinder: a multilinguistic token-based code
  clone detection system for large scale source code, IEEE Transactions on
  Software Engineering 28~(7) (2002) 654--670.

\bibitem{jiang2007deckard}
L.~Jiang, G.~Misherghi, Z.~Su, S.~Glondu, Deckard: Scalable and accurate
  tree-based detection of code clones, in: Proceedings of the 29th
  international conference on Software Engineering, IEEE Computer Society,
  2007, pp. 96--105.

\bibitem{baxter2004dms}
I.~D. Baxter, C.~Pidgeon, M.~Mehlich, Dms/spl reg: program transformations for
  practical scalable software evolution, in: Software Engineering, 2004. ICSE
  2004. Proceedings. 26th International Conference on, IEEE, 2004, pp.
  625--634.

\bibitem{basit2005detecting}
H.~A. Basit, S.~Jarzabek, Detecting higher-level similarity patterns in
  programs, in: ACM Sigsoft Software engineering notes, Vol.~30, ACM, 2005, pp.
  156--165.

\bibitem{chen2017damgate}
Y.~Chen, T.~Lan, G.~Venkataramani, Damgate: dynamic adaptive multi-feature
  gating in program binaries, in: Proceedings of the 2017 Workshop on Forming
  an Ecosystem Around Software Transformation, ACM, 2017, pp. 23--29.

\bibitem{chen2018toss}
Y.~Chen, S.~Sun, T.~Lan, G.~Venkataramani, Toss: Tailoring online server
  systems through binary feature customization, in: Proceedings of the 2018
  Workshop on Forming an Ecosystem Around Software Transformation, ACM, 2018,
  pp. 1--7.

\bibitem{caballero2012undangle}
J.~Caballero, G.~Grieco, M.~Marron, A.~Nappa, Undangle: early detection of
  dangling pointers in use-after-free and double-free vulnerabilities, in:
  Proceedings of the 2012 International Symposium on Software Testing and
  Analysis, ACM, 2012, pp. 133--143.

\bibitem{serna2012info}
F.~J. Serna, The info leak era on software exploitation, Black Hat USA.

\bibitem{conti2015losing}
M.~Conti, S.~Crane, L.~Davi, M.~Franz, P.~Larsen, M.~Negro, C.~Liebchen,
  M.~Qunaibit, A.-R. Sadeghi, Losing control: On the effectiveness of
  control-flow integrity under stack attacks, in: Proceedings of the 22nd ACM
  SIGSAC Conference on Computer and Communications Security, ACM, 2015, pp.
  952--963.

\bibitem{sajnani2016sourcerercc}
H.~Sajnani, V.~Saini, J.~Svajlenko, C.~K. Roy, C.~V. Lopes, Sourcerercc:
  Scaling code clone detection to big-code, in: Software Engineering (ICSE),
  2016 IEEE/ACM 38th International Conference on, IEEE, 2016, pp. 1157--1168.

\bibitem{allamanis2017survey}
M.~Allamanis, E.~T. Barr, P.~Devanbu, C.~Sutton, A survey of machine learning
  for big code and naturalness, arXiv preprint arXiv:1709.06182.

\bibitem{joern}
F.~Yamaguchi, {Joern: A Robust Code Analysis Platform for C/C++},
  \url{http://www.mlsec.org/joern/} (2016).

\bibitem{cadar2008klee}
C.~Cadar, D.~Dunbar, D.~R. Engler, et~al., Klee: Unassisted and automatic
  generation of high-coverage tests for complex systems programs., in: OSDI,
  Vol.~8, 2008, pp. 209--224.

\bibitem{links}
{Twibright Labs}, Links, \url{http://links.twibright.com}.

\bibitem{thttpd_ACME}
{ACME Lab}, Thttpd, \url{http://www.acme.com/software/thttpd/}.

\bibitem{milea2014vector}
N.~A. Milea, L.~Jiang, S.-C. Khoo, Vector abstraction and concretization for
  scalable detection of refactorings, in: Proceedings of the 22nd ACM SIGSOFT
  International Symposium on Foundations of Software Engineering, ACM, 2014,
  pp. 86--97.

\bibitem{baker1993program}
B.~S. Baker, A program for identifying duplicated code, Computing Science and
  Statistics (1993) 49--49.

\bibitem{parr1995antlr}
T.~J. Parr, R.~W. Quong, Antlr: A predicated-ll (k) parser generator, Software:
  Practice and Experience 25~(7) (1995) 789--810.

\bibitem{baxter1998clone}
I.~D. Baxter, A.~Yahin, L.~Moura, M.~Sant'Anna, L.~Bier, Clone detection using
  abstract syntax trees, in: Software Maintenance, 1998. Proceedings.,
  International Conference on, IEEE, 1998, pp. 368--377.

\bibitem{datar2004locality}
M.~Datar, N.~Immorlica, P.~Indyk, V.~S. Mirrokni, Locality-sensitive hashing
  scheme based on p-stable distributions, in: Proceedings of the twentieth
  annual symposium on Computational geometry, ACM, 2004, pp. 253--262.

\bibitem{gionis1999similarity}
A.~Gionis, P.~Indyk, R.~Motwani, et~al., Similarity search in high dimensions
  via hashing, in: Vldb, Vol.~99, 1999, pp. 518--529.

\bibitem{zheng2013z3}
Y.~Zheng, X.~Zhang, V.~Ganesh, Z3-str: A z3-based string solver for web
  application analysis, in: Proceedings of the 2013 9th Joint Meeting on
  Foundations of Software Engineering, ACM, 2013, pp. 114--124.

\bibitem{brummayer2009boolector}
R.~Brummayer, A.~Biere, Boolector: An efficient smt solver for bit-vectors and
  arrays, in: International Conference on Tools and Algorithms for the
  Construction and Analysis of Systems, Springer, 2009, pp. 174--177.

\bibitem{veanes2010symbolic}
M.~Veanes, N.~Bj{\o}rner, L.~De~Moura, Symbolic automata constraint solving,
  in: International Conference on Logic for Programming Artificial Intelligence
  and Reasoning, Springer, 2010, pp. 640--654.

\bibitem{webber2012programmatic}
J.~Webber, A programmatic introduction to neo4j, in: Proceedings of the 3rd
  annual conference on Systems, programming, and applications: software for
  humanity, ACM, 2012, pp. 217--218.

\bibitem{spec}
{SPEC CPU 2006}, \url{https://www.spec.org/cpu2006/} (2006).

\bibitem{lu2005bugbench}
S.~Lu, Z.~Li, F.~Qin, L.~Tan, P.~Zhou, Y.~Zhou, Bugbench: Benchmarks for
  evaluating bug detection tools, in: Workshop on the evaluation of software
  defect detection tools, Vol.~5, 2005.

\bibitem{libreoffice}
LibreOffice, Cve-2018-10120,
  \url{https://cve.mitre.org/cgi-bin/cvename.cgi?name=CVE-2018-10120} (2018).

\bibitem{ducasse1999language}
S.~Ducasse, M.~Rieger, S.~Demeyer, A language independent approach for
  detecting duplicated code, in: Software Maintenance, 1999.(ICSM'99)
  Proceedings. IEEE International Conference on, IEEE, 1999, pp. 109--118.

\bibitem{baker1995finding}
B.~S. Baker, On finding duplication and near-duplication in large software
  systems, in: Reverse Engineering, 1995., Proceedings of 2nd Working
  Conference on, IEEE, 1995, pp. 86--95.

\bibitem{kontogiannis1996pattern}
K.~A. Kontogiannis, R.~DeMori, E.~Merlo, M.~Galler, M.~Bernstein, Pattern
  matching for clone and concept detection, Automated Software Engineering
  3~(1-2) (1996) 77--108.

\bibitem{wahler2004clone}
V.~Wahler, D.~Seipel, J.~Wolff, G.~Fischer, Clone detection in source code by
  frequent itemset techniques, in: Source Code Analysis and Manipulation, 2004.
  Fourth IEEE International Workshop on, IEEE, 2004, pp. 128--135.

\bibitem{zhang2014personal}
K.~Zhang, M.~Wang, X.~Cong, F.~Huang, H.~Xue, L.~Li, Z.~Gao, Personal
  attributes extraction based on the combination of trigger words, dictionary
  and rules, in: Proceedings of The Third CIPS-SIGHAN Joint Conference on
  Chinese Language Processing, 2014, pp. 114--119.

\bibitem{emami1994context}
M.~Emami, R.~Ghiya, L.~J. Hendren, Context-sensitive interprocedural points-to
  analysis in the presence of function pointers, in: ACM SIGPLAN Notices,
  Vol.~29, ACM, 1994, pp. 242--256.

\bibitem{heintze2001demand}
N.~Heintze, O.~Tardieu, Demand-driven pointer analysis, in: ACM SIGPLAN
  Notices, Vol.~36, ACM, 2001, pp. 24--34.

\bibitem{nystrom2004bottom}
E.~M. Nystrom, H.-S. Kim, W.~H. Wen-mei, Bottom-up and top-down
  context-sensitive summary-based pointer analysis, in: International Static
  Analysis Symposium, Springer, 2004, pp. 165--180.

\bibitem{avgerinos2014automatic}
T.~Avgerinos, S.~K. Cha, A.~Rebert, E.~J. Schwartz, M.~Woo, D.~Brumley,
  Automatic exploit generation, Communications of the ACM 57~(2) (2014) 74--84.

\bibitem{godefroid2012sage}
P.~Godefroid, M.~Y. Levin, D.~Molnar, Sage: whitebox fuzzing for security
  testing, Communications of the ACM 55~(3) (2012) 40--44.

\bibitem{hu2017binary}
Y.~Hu, Y.~Zhang, J.~Li, D.~Gu, Binary code clone detection across architectures
  and compiling configurations, in: Program Comprehension (ICPC), 2017 IEEE/ACM
  25th International Conference on, IEEE, 2017, pp. 88--98.

\bibitem{xue2020learn2reason}
H.~Xue, Learn2reason: Joint statistical and formal learning approach to improve
  the robustness and time-to-solution for software security, Ph.D. thesis, The
  George Washington University (2020).

\bibitem{pewny2015cross}
J.~Pewny, B.~Garmany, R.~Gawlik, C.~Rossow, T.~Holz, Cross-architecture bug
  search in binary executables, in: Security and Privacy (SP), 2015 IEEE
  Symposium on, IEEE, 2015, pp. 709--724.

\bibitem{xue2018clone}
H.~Xue, G.~Venkataramani, T.~Lan, Clone-slicer: Detecting domain specific
  binary code clones through program slicing, in: Proceedings of the 2018
  Workshop on Forming an Ecosystem Around Software Transformation, ACM, 2018,
  pp. 27--33.

\bibitem{xue2020twin}
H.~Xue, Y.~Mei, K.~Gogineni, G.~Venkataramani, T.~Lan, Twin-finder: Integrated
  reasoning engine for pointer-related code clone detection, in: 2020 IEEE 14th
  International Workshop on Software Clones (IWSC), IEEE, 2020, pp. 1--7.

\bibitem{xue2018clone_hunter}
H.~Xue, G.~Venkataramani, T.~Lan, Clone-hunter: accelerated bound checks
  elimination via binary code clone detection, in: Proceedings of the 2nd ACM
  SIGPLAN International Workshop on Machine Learning and Programming Languages,
  ACM, 2018, pp. 11--19.

\bibitem{xue2018morph}
H.~Xue, Y.~Chen, G.~Venkataramani, T.~Lan, G.~Jin, J.~Li, Morph: Enhancing
  system security through interactive customization of application and
  communication protocol features, in: Proceedings of the 2018 ACM SIGSAC
  Conference on Computer and Communications Security, ACM, 2018, pp.
  2315--2317.

\bibitem{xue2019hecate}
H.~Xue, Y.~Chen, G.~Venkataramani, T.~Lan, Hecate: Automated customization of
  program and communication features to reduce attack surfaces, in:
  International Conference on Security and Privacy in Communication Systems,
  Springer, 2019.

\bibitem{chen2020chop}
Y.~Chen, H.~Xue, T.~Lan, G.~Venkataramani, Chop: Bypassing runtime bounds
  checking through convex hull optimization, Computers \& Security 90 (2020)
  101708.

\bibitem{yao2017statsym}
F.~Yao, Y.~Li, Y.~Chen, H.~Xue, T.~Lan, G.~Venkataramani, Statsym: vulnerable
  path discovery through statistics-guided symbolic execution, in: Dependable
  Systems and Networks (DSN), 2017 47th Annual IEEE/IFIP International
  Conference on, IEEE, 2017, pp. 109--120.

\bibitem{li2016sarre}
Y.~Li, F.~Yao, T.~Lan, G.~Venkataramani, Sarre: semantics-aware rule
  recommendation and enforcement for event paths on android, IEEE Transactions
  on Information Forensics and Security 11~(12) (2016) 2748--2762.

\bibitem{xue2017simber}
H.~Xue, Y.~Chen, F.~Yao, Y.~Li, T.~Lan, G.~Venkataramani, Simber: Eliminating
  redundant memory bound checks via statistical inference, in: IFIP
  International Conference on ICT Systems Security and Privacy Protection,
  Springer, 2017, pp. 413--426.

\bibitem{yamaguchi2013chucky}
F.~Yamaguchi, C.~Wressnegger, H.~Gascon, K.~Rieck, Chucky: Exposing missing
  checks in source code for vulnerability discovery, in: Proceedings of the
  2013 ACM SIGSAC conference on Computer \& communications security, ACM, 2013,
  pp. 499--510.

\end{thebibliography}

\end{document}